\documentclass[12pt, a4paper]{article}
\pdfoutput=1

\usepackage[utf8]{inputenc} 
\usepackage[T1]{fontenc}

\usepackage{amsmath}
\usepackage{amssymb}
\usepackage{amsfonts}
\usepackage{bbm}
\usepackage{verbatim}
\usepackage{dsfont}
\usepackage{booktabs}
\usepackage{slashed}
\usepackage{bm}
\usepackage{subcaption}
\usepackage{cancel}

\usepackage{mathtools}

\usepackage{epsfig}
\usepackage{color}
\usepackage[table]{xcolor}
\usepackage{graphicx}
\usepackage[]{caption}
\usepackage{float}
\usepackage{overpic}

\usepackage{enumerate}
\usepackage{hhline}
\usepackage{multirow}
\usepackage{xspace}
\usepackage{setspace}
\usepackage[title]{appendix}

\renewcommand{\theequation}{\arabic{section}.\arabic{equation}}

\newcommand{\F}{{\cal F}}

\newcommand{\V}{{\cal V}}
\renewcommand{\S}{{\cal S}}
\newcommand{\cH}{{\cal H}}

\newcommand{\Z}{\mathbb{Z}}

\renewcommand{\Re}{{\rm Re}\,}
\renewcommand{\Im}{{\rm Im}\,}

\newcommand{\ie}{{\em i.e.} }

\renewcommand{\and}{\mbox{and}}

\usepackage{xcolor}

\newcommand{\Nf}{N_{\rm flux}}

\newcommand{\rank}{{\rm rank}\,}

\newcommand{\half}{\frac{1}{2}} 

\renewcommand{\bm}{\boldmath}

\def\be{\begin{equation}}
\def\ee{\end{equation}}
\def\bea{\begin{eqnarray}}
\def\eea{\end{eqnarray}}
\def\bes{\begin{subequations}}
\def\ees{\end{subequations}}

\renewcommand{\a}{\alpha}

\usepackage[square,numbers,compress]{natbib}
\usepackage{bibentry}

\counterwithin*{equation}{section}

\usepackage[hidelinks, linkcolor=blue, citecolor=blue, urlcolor=blue]{hyperref}

\begin{document}

\thispagestyle{empty}

\begin{flushright}
{IFT-UAM/CSIC-22-148}\\
\end{flushright}
\vskip .8 cm
\begin{center}
  {\Large {\bf 
 Analytics of type IIB flux vacua \\[7pt] and their mass spectra
}}\\[12pt]

\bigskip
\bigskip 
\bigskip
{\bf Thibaut Coudarchet,}\footnote{thibaut.coudarchet@uam.es}\textsuperscript{,*}
{\bf Fernando Marchesano,}\footnote{fernando.marchesano@csic.es}\textsuperscript{,*}
{\bf David Prieto}\footnote{david.prietor@estudiante.uam.es}\textsuperscript{,*}\\
{\bf and Mikel A. Urkiola}\footnote{mikel.alvarezu@ehu.eus}\textsuperscript{,*,$\dagger$}

\setcounter{footnote}{0}

\bigskip
\bigskip
\vspace{0.23cm}
\textsuperscript{*}{\it Instituto de F\'isica Te\'orica UAM-CSIC, c/ Nicol\'as Cabrera 13-15,\\ 28049 Madrid, Spain}\\[5pt] 
\textsuperscript{$\dagger$}{\it Department of Applied Mathematics, University of the Basque Country UPV/EHU, 48013 Bilbao, Spain}\\[20pt] 
\bigskip
\end{center}

\date{\today{}}

\begin{abstract}
\noindent
We analyze the tree-level potential of type IIB flux compactifications in warped Calabi--Yau orientifolds, in regions of weak coupling and moderately large complex structure. In this regime, one may approximate the flux-induced superpotential $W$ by a polynomial on the axio-dilaton and complex structure fields, and a significant fraction of vacua corresponds to a quadratic $W$. In this quadratic case, we argue that vacua fall into three classes, for which one can push the analytic description of their features. In particular, we provide analytic expressions for the vacuum expectation values and flux-induced masses of the axio-dilaton and complex structure fields in a large subclass of vacua, independently of the Calabi--Yau and the number of moduli. We show that supersymmetric vacua always contain flat directions, at least at this level of approximation. Our findings allow to generate vast ensembles of flux vacua in specific Calabi--Yau geometries, as we illustrate in a particular example. 

\end{abstract}

\newpage 
\setcounter{page}{2}

\renewcommand{\baselinestretch}{1.5}


\tableofcontents

\section{Introduction}
\label{sec:Introduction}

Our current picture of the string Landscape is tightly connected to the different mechanisms for moduli stabilization. This is because a simple procedure to generate an ensemble of vacua is to consider an effective field theory (EFT) with a perturbative multi-dimensional moduli space ${\cal M}$, and implement one or several moduli-fixing mechanisms that select a discrete set of points in ${\cal M}$. In string theory compactifications, this philosophy can be realized by means of background fluxes threading the internal dimensions  \cite{Grana:2005jc,Douglas:2006es,Blumenhagen:2006ci,Becker:2006dvp,Marchesano:2007de,Denef:2008wq,Denef:2007pq,Ibanez:2012zz,Quevedo:2014xia,Baumann:2014nda}, so that the discretum of vacua is a consequence of flux quantization. Particularly simple is the case of type IIB Calabi--Yau  (CY) orientifolds with three-form fluxes. In the absence of strongly warped regions \cite{Giddings:2001yu}, the main effect of these fluxes is to generate a superpotential for the axio-dilaton and complex structure fields \cite{Gukov:1999ya}.  Thus, from a single Calabi--Yau geometry and below the scale of flux-induced masses, one obtains an ensemble of 4d EFTs indexed by the three-form flux quanta,\footnote{Note that a 4d EFT of this sort has fixed NS three-form flux quanta \cite{Lanza:2019xxg}, so there is an ensemble of 4d EFTs even at the scale of flux-induced masses, giving rise to a larger ensemble at lower scales.} whose physics can be extracted from the same parent (fluxless) 4d EFT.

Despite its relative simplicity, in practice there is not much analytic control when describing this setup. In particular, as soon as there are several complex structure moduli stabilized by fluxes, the analytic description of the set of vacua is typically lost, except in some special cases where the use of discrete isometry groups allows for a consistent reduction of the complex structure sector \cite{Kachru:2003aw,Balasubramanian:2005zx,Conlon:2005ki,
Westphal:2006tn,Giryavets:2003vd,Giryavets:2004zr,DeWolfe:2004ns,Denef:2004dm,Louis:2012nb,Cicoli:2013cha} possibly down to a single field \cite{Klemm:1992tx,Doran:2007jw,Candelas:2017ive,Braun:2011hd,Batyrev:2008rp,Doran:2005gu,Candelas:2019llw,Joshi:2019nzi,Grimm:2019ixq,Blanco-Pillado:2020wjn}. The same statement applies to the mass spectrum of the fields that are stabilized by fluxes, which depends on the scalar potential and the vacuum expectation values (vevs) of the fields. These two ingredients, vevs and mass spectra, are crucial in order to implement full moduli stabilization, and therefore to develop an overall picture of the ensemble of vacua and to extract its phenomenological features. 

The aim of this paper is to improve the current state of affairs, by providing a class of type IIB flux configurations where the vevs and mass spectrum in the axio-dilaton and complex structure sector can be described analytically.\footnote{In most type IIB CY schemes that implement full moduli stabilization, the flux-induced vevs and masses are independent of the K\"ahler moduli stabilization details, and can therefore be seen as properties of the final vacuum. In this paper we will not discuss K\"ahler moduli stabilization, and we will dub as {\em flux vacua} those vevs in the axio-dilaton and complex structure sector that solve their equations of motion at tree-level in 4d Minkowski.} This analytic description is independent of the number of complex structure fields, and the key ingredient to implement it is a simplified description of the Calabi--Yau holomorphic three-form periods in some asymptotic region. We focus on the region of Large Complex Structure (LCS), where such periods can be expressed as polynomials of the complex structure fields, up to exponential terms that can be neglected. It is precisely in this region where recent progress in describing the flux-induced mass spectrum \cite{Blanco-Pillado:2020wjn,Blanco-Pillado:2020hbw} and the flux potential \cite{Marchesano:2021gyv} analytically and for an arbitrary number of fields has been made, so it is a particularly promising regime to look at. In this work we show how these two different set of results are connected to each other, and how they can be merged into a single framework that leads to a more detailed analysis of such flux vacua. 

Indeed, as pointed out in \cite{Marchesano:2021gyv,Sousa:2014qza,Marsh:2015zoa,Brodie:2015kza}, in order to find vacua in the LCS limit, the flux contribution to the D3-brane tadpole must grow with the field vevs, unless certain flux quanta are set to zero. A particular family of flux configurations avoiding this problem was proposed in \cite{Marchesano:2021gyv}, and dubbed \emph{IIB1 scenario} therein. As we show in this work, this family corresponds to a set of compactifications in which the flux-induced superpotential is quadratic in the axio-dilaton and complex structure fields. It follows from here that the set of flux vacua splits into three distinct classes, that can be classified according to the nature of the field directions that are unfixed by fluxes.\footnote{More precisely, these are flat directions at the approximation level in which all polynomial corrections to the leading behaviour of the periods are included, while exponential corrections are neglected.} In the first class, in which supersymmetry is broken in the K\"ahler sector, all fields in the complex structure/axio-dilaton sector are stabilized. Moreover, the simplest choice of fluxes leads to the {\em no-scale aligned} vacua of \cite{Blanco-Pillado:2020wjn}. In this case, one can describe the field vevs in terms of quadratic and cubic equations, and apply the techniques of \cite{Blanco-Pillado:2020wjn} to obtain the flux-induced mass spectrum analytically, for an arbitrary number of complex structure moduli. The second class also breaks supersymmetry in the K\"ahler sector, but now contains one or more axion-like fields that are flat directions of the flux potential. Finally, in the third class, vacua are fully supersymmetric and, remarkably, they always contain some complexified flat directions.

These results can be compared to other strategies in the literature employed to analyze the same setup. For instance, one may compute the flux-induced mass spectrum by first extracting the Hessian from the analytic expression for the scalar potential provided in \cite{Marchesano:2021gyv}. While this analysis is in general quite involved, one can see that for the axionic sector of the IIB1 scenario one obtains a perfect match with our analytic expressions. A different, more direct method is to perform a numerical analysis of the flux vacua solutions and their mass spectra. When applying this approach to the IIB1 scenario the result is two-fold: On the one hand, it shows that the analytical control inside the IIB1 setup allows to very efficiently find flux configurations yielding consistent vacua. On the other hand, various features of the numerical vacua are shown to precisely match the analytical results developed in the paper, supporting the robustness of the analysis.

The paper is organized as follows: In sect.~\ref{sec:Notations} we define usual notations and conventions for type IIB flux compactifications at LCS. In sect.~\ref{sec:Bilinear}, we provide a coarse-grained classification of vacua that can arise from a quadratic superpotential and uncover the supersymmetric and the two non-supersymmetric families mentioned above. We detail here what is the \emph{IIB1 scenario} for which, precisely, the superpotential takes a bilinear form. In sect.~\ref{sec:nonsusy}, we explore the non-supersymmetric vacua highlighted in the generic classification in more detail. We focus on a specific branch of vacua by assuming an ansatz for the saxions, where, upon further refinement to two cases, we can express analytically the vacuum expectation values of the axio-dilaton  and all complex structure moduli. We prove here that one of these two cases falls into the \emph{no-scale aligned} class described in \cite{Blanco-Pillado:2020wjn}, so that we are able to determine their complete tree-level mass spectra analytically. Details about the computation of these masses are presented in appendices \ref{sec:NSA} and \ref{ap:spot}. In sect.~\ref{sec:susy}, we briefly investigate the supersymmetric family exhibited from the generic classification. In sect.~\ref{sec:numerics}, we numerically generate and analyze an ensemble of IIB1 vacua that fits into the \emph{no-scale aligned} branch in a toy two-parameter model. We end up  with some conclusions and prospects in sect.~\ref{sec:Conclusion}.


\section{Generics of type IIB flux compactifications}
\label{sec:Notations}

In this section, we review some usual definitions and notations about the effective supergravity of type IIB string theory compactified on a Calabi--Yau 3-fold $X_3$.

\subsection{The prepotential}

In a symplectic basis $\{A^I,B_I\}$, $I=0,\dots,h^{2,1}$ of $H_3(X_3,\Z)$, the periods of the Calabi--Yau $(3,0)$-form $\Omega$ are encoded in the vector
\begin{equation}
\Pi^t\equiv(\F_I,X^I)=\left(\int_{B_I}\Omega,\int_{A^I}\Omega\right)\ ,
\end{equation}
where $t$ stands for the transpose. The complex structure moduli fields are defined to be $z^i\equiv X^i/X^0$, $i=1,\dots,h^{2,1}$ and the $\F_I$ components are expressed as derivatives of the \emph{prepotential} $\F$. Setting the gauge $X^0=1$, the period vector takes the following form:
\begin{equation}
\label{eq:period}
    \Pi = 
    \begin{pmatrix}
     2\mathcal{F} - z^i \partial_i \F \\
    \partial_i \F \\
     1 \\
    z^i 
    \end{pmatrix}.
\end{equation}

In the LCS regime the prepotential reads
\begin{equation}
\F=-\frac{1}{6}\kappa_{ijk}z^iz^jz^k-\half a_{ij}z^iz^j+c_iz^i+\half\kappa_0+\F_{\rm inst}\ .
\label{eq:full_prepotential}
\end{equation}
The instanton contribution $\F_{\rm inst}$ is subleading in the LCS regime and can be expressed as sum of polylogarithm $\text{Li}_p(q)\equiv \sum_{k>0}\frac{q^k}{k^p}$ ponderated by Gopakumar-Vafa invariants $n_{\vec d}$ labeled by $\vec d\in(\mathbb{Z}^+)^{h^{2,1}}$ \cite{Cicoli:2013cha},
\begin{equation}
\F_{\rm inst}=-\frac{i}{(2\pi)^3}\sum_{\vec d}n_{\vec d}\, \text{Li}_3[e^{-2\pi d_iz^i}]\ .
\end{equation}
The coefficients $\kappa_{ijk}$, $a_{ij}$ and $c_i$ can be computed from topological data of the mirror manifold $Y_3$ of the Calabi--Yau $X_3$, while $\kappa_0$ depends on the Euler characteristic of $X_3$. More precisely, we have \cite{Mayr:2000as}
\begin{align}
\begin{split}
&\kappa_{ijk}\equiv\int_{Y_3}\omega_i\wedge\omega_j\wedge\omega_k\ ,\qquad a_{ij}\equiv-\half\int_{Y_3}\omega_i\wedge i_*\text{ch}_1(\text{P.D}[w_j])\ ,\\
&c_i\equiv\frac{1}{24}\int_{Y_3}\omega_i\wedge \text{ch}_2(Y_3)\ ,\qquad \kappa_0\equiv\frac{\zeta(3)\chi(X_3)}{(2\pi i)^3}\ = i \, \frac{\zeta (3)}{4\pi^3} (h^{1,1} - h^{2,1})\ ,
\end{split}
\end{align}
where $\omega_i$, $i=1,\dots,h^{1,1}(Y_3)$ form a basis of $H^2(Y_3,\Z)$, $i_*$ denotes the pushforward of the embedding $i$ of the divisors into $Y_3$, P.D stands for Poincaré Dual and $\text{ch}_1$ and $\text{ch}_2$ denote the first and second Chern classes respectively. It can further be shown \cite{Cicoli:2013cha} that $a_{ij}$ can be rewritten in terms of the triple intersection numbers as follows
\begin{equation}
    a_{ij} = -\frac{1}{2} \int_{Y_3} \omega_i \wedge \omega_j \wedge \omega_j  \mod \mathbb{Z}\ .
\end{equation}
Finally, it is important to note that both $c_i$ and $a_{ij}$ are defined only modulo $\mathbb{Z}$, since shifts on these parameters correspond to different choices for the symplectic basis of 3-cycles of $X_3$. This leads to important restrictions on their values, when considered in terms of the transformation properties of the period vector under monodromies $z^i\to z^i+v^i$, $v^i\in\Z$ at LCS. More concretely, the coefficients of the prepotential must satisfy the following conditions \cite{Mayr:2000as}: 
\begin{align}
    a_{ij} + \frac{1}{2} \kappa_{ijj} \in \mathbb{Z}\quad\text{ and }\quad
    2c_i + \frac{1}{6} \kappa_{iii} \in \mathbb{Z}\label{consistency}\ .
\end{align}
The first equation can also be generalized to take the form
\begin{equation}
\label{freedom}
a_{ij}v^j+\half\kappa_{ijk}v^jv^k= 0 \!\!\!\mod \mathbb{Z}\ .
\end{equation}
Note that we can make use of the redundancy of $a_{ij}$ to shift its value like $a_{ij} \to a_{ij} + n_{ij}$, $n_{ij} \in \mathbb{Z}$ so that the LHS of \eqref{freedom} is actually 0.

\subsection{Kähler potential}

The tree-level Kähler potential is given by 
\begin{equation}
K= K_{\text{k}} + K_{\text{dil}} + K_{\text{cs}} = -2\log(\V)-\log(-i(\tau-\bar\tau))-\log(-i\Pi^\dagger\cdot\Sigma\cdot\Pi)\ ,
\end{equation}
where $\V$ is the volume of $X_3$, $\tau$ is the axio-dilaton and we have defined the canonical symplectic $(2h^{2,1}+2)\times(2h^{2,1}+2)$ matrix
\begin{equation}
\Sigma\equiv\begin{pmatrix}
0 & \mathds{1}\\ -\mathds{1} & 0
\end{pmatrix}.
\end{equation}
The Kähler potential at the approximation of large complex structure can be shown to read
\begin{align}
    K_{\text{cs}} &= - \log \left( \frac{i}{6} \kappa_{ijk} (z^i-\bar{z}^i) (z^j-\bar{z}^j) (z^k-\bar{z}^k) - 2 \, \Im (\kappa_0) \right) \nonumber \\
    &= - \log \left( \frac{4}{3} \kappa_{ijk} t^i t^j t^k - 2 \, \Im (\kappa_0) \right)\ ,
\label{eq: complex Kahler potential}
\end{align}
where we have defined $z^i \equiv b^i + i t^i$ and, for later use, we also introduce $\tau \equiv b^0 + i t^0$. 

It will be important to develop some of the derivatives of the Kähler potential, for future reference. The most relevant ones are the following:
\begin{align}
    K_{\tau} &= - \frac{1}{\tau - \bar{\tau}} = \frac{i}{2 t^0}\ , \label{eq:Kt}\\
    K_{\tau \bar{\tau}} &= - \frac{1}{(\tau - \bar{\tau})^2} = \frac{1}{4 (t^0)^2}\ , \\
    K_i &= - \frac{i}{2} \mathring{\kappa}_{ijk} (z^j - \bar{z}^j) (z^k - \bar{z}^k) = 2 i \mathring{\kappa}_{ijk} t^j t^k\ , \label{eq:Ki}\\
    K_{i\bar{j}} &= i \mathring{\kappa}_{ijk} (z^k - \bar{z}^k) + \frac{1}{4} \mathring{\kappa}_{imn} \mathring{\kappa}_{jpq} (z^m - \bar{z}^m) (z^n - \bar{z}^n)  (z^p - \bar{z}^p) (z^q - \bar{z}^q) \nonumber \\
    &= - 2 \mathring{\kappa}_{ijk} t^k + 4 \mathring{\kappa}_{imn} \mathring{\kappa}_{jpq} t^m t^n t^p t^q\ ,\label{eq:Kij}
\end{align}
where we have defined $\mathring{\kappa}_{ijk} \equiv e^{K_{\rm cs}} \kappa_{ijk}$ and the indices $\tau$ and $i$ denote derivatives of the Kähler potential with respect to the axio-dilaton and the complex structure moduli $z^i$ respectively (barred indices naturally denote derivatives with respect to the complex conjugate fields).

Intuitively, the LCS regime establishes how the cubic term inside the previous logarithm compares with the constant contribution $\kappa_0$. Thus, we introduce the following \emph{LCS parameter} to measure how close to the LCS point a given solution is:
\begin{align}
\label{eq:def_xi}
    \xi \equiv \frac{- 2 \, \Im (\kappa_0)}{\frac{4}{3} \kappa_{ijk} t^i t^j t^k} = \frac{-2 e^{K_{\text{cs}}} \Im (\kappa_0)}{1 + 2 e^{K_{\text{cs}}} \Im (\kappa_0)}\ .
\end{align}
By definition, the LCS point is located at $\xi = 0$. On the other hand, it can be checked that in those geometries where $h^{2,1} > h^{1,1}$, we obtain negative eigenvalues in the field-space metric $K_{i\bar{j}}$ if $\xi > 1/2$, thus rendering those solutions unphysical; As for geometries with $h^{2,1} > h^{1,1}$, solutions with $\xi<-1$ will suffer from the same problem \cite{Blanco-Pillado:2020wjn}.

\subsection{Flux superpotential}

With these definitions, we can express the usual Gukov-Vafa-Witten (GVW) superpotential $W$ \cite{Gukov:1999ya}, induced by fluxes threading the compact geometry. 
We first introduce the flux vector
\begin{equation}
N\equiv f-\tau h\ \ \ \ \text{with}\ \ \ \ 
f\equiv\begin{pmatrix}
\int_{B^I} F_3 \\ 
\int_{A_I} F_3
\end{pmatrix} 
\equiv
\begin{pmatrix}
f^B_0 \\ f^B_i \\ f_A^0 \\ f_A^i 
\end{pmatrix}
\ \ \ \ \text{and}\ \ \ \
h\equiv\begin{pmatrix}
\int_{B^I}H_3\\
\int_{A_I}H_3
\end{pmatrix}
\equiv
\begin{pmatrix}
h^B_0 \\ h^B_i \\ h_A^0 \\ h_A^i 
\end{pmatrix}.
\end{equation}
These fluxes induce a D3-tadpole Ramond-Ramond charge in the compact space, which has to be cancelled by negatively charged objects, like orientifold planes. The full D3-charge $\Nf$ induced by these fluxes is shown to be
\begin{equation}
    \Nf = f^T \cdot \Sigma \cdot h = - \frac{N^{\dag}\cdot \Sigma \cdot N}{\tau - \bar{\tau}}\ .
\end{equation}

The GVW superpotential can then be easily expressed as\footnote{Note that we deliberately forget a factor $1/\sqrt{4\pi}$ since it will be irrelevant for the vacuum equations and everything we will compute.} \cite{Gukov:1999ya}
\begin{equation}
W\equiv\int (F_3-\tau H_3)\wedge \Omega=N^T\cdot\Sigma\cdot\Pi\ .
\end{equation}
From this equation we can obtain the full expression for the superpotential, which reads
\begin{align}
\begin{split}
W = &- \frac{1}{6} N_A^0 \kappa_{ijk} z^i z^j z^k + \frac{1}{2} \kappa_{ijk} N_A^i z^j z^k + \left( N_A^j a_{ij} + N_i^B - N_A^0 c_i \right) z^i\\
&- \kappa_0 N_A^0 - N_A^i c_i + N_0^B\ .
\label{eq:Wfull}
\end{split}
\end{align}

\subsection{Vacuum equations}

At tree-level, type IIB Calabi--Yau compactifications with three-form fluxes yield 4d Minkowski vacua.  Since the 4d EFT features a no-scale structure in the Kähler sector ($K^{\rho\sigma}K_\rho K_\sigma=3$ where $\rho, \sigma$ run over Kähler moduli), the corresponding vacua equations are given by $D_A W\equiv\partial_AW+K_AW= 0, \ A\in\left\lbrace \tau,z^i \right\rbrace$. Let us write these equations explicitly:
\begin{align}
D_\tau W &= \left[ - h  - \frac{1}{\tau - \bar{\tau}} (f - \tau h) \right]^T \cdot \Sigma \cdot \Pi = - \frac{1}{\tau - \bar{\tau}} \bar{N}^T \cdot \Sigma \cdot \Pi = 0\ , \\[5pt]
D_i W &= N^T \cdot \Sigma \cdot D_i \Pi = 0\ ,
\end{align}
which translate into
\begin{align}
\label{eq:DtW_DiW_full}
& - \frac{1}{6} \bar{N}_A^0 \kappa_{ijk} z^i z^j z^k + \frac{1}{2} \kappa_{ijk} \bar{N}_A^i z^j z^k + \left( \bar{N}_A^j a_{ij} + \bar{N}_i^B - \bar{N}_A^0 c_i \right) z^i - \kappa_0 \bar{N}_A^0 - \bar{N}_A^i c_i + \bar{N}_0^B = 0\ , \nonumber\\[5pt]
& - \frac{1}{2} N_A^0 \kappa_{ijk} z^j z^k + \kappa_{ijk} N_A^j z^k + \left( N_A^j a_{ij} + N_i^B - N_A^0 c_i \right) + K_i W = 0\ .
\end{align}
Supersymmetric vacua are realized if, in addition, the covariant derivatives of the superpotential with respect to the Kähler moduli are zero. Since they are proportional to $W$, the superpotential should vanish to yield a supersymmetric vacuum. Namely, with $\sigma$ referring to the Kähler sector:
\begin{equation}
\text{Supersymmetric condition: }D_\sigma W=K_\sigma W=0 \Longleftrightarrow W=0\ .
\end{equation}

\subsection{Various contractions with triple intersection numbers}

We define here several notations we use in the paper to describe the triple intersection number $\kappa_{ijk}$ contracted with various quantities. They will be redefined in the sequel at the appropriate moment but we find useful to have them summarized here. We denote
\begin{equation}
\begin{aligned}
&\kappa_{ij}\equiv \kappa_{ijk}t^k\ , &&\kappa_{i}\equiv\kappa_{ijk}t^jt^k\ , &&\kappa\equiv\kappa_{ijk}t^it^jt^k\ ,\\
&S_{ij}\equiv \kappa_{ijk}f_A^k\ ,&&S_i\equiv\kappa_{ijk}f_A^jf_A^k\ ,&&\S\equiv\kappa_{ijk}f_A^if_A^jf_A^k\ ,\\
&\kappa_{ij}^\cH\equiv \kappa_{ijk}S^{kn}h_n^B\ ,\ \ &&\kappa_i^\cH\equiv\kappa_{ijk}S^{jm}h_m^BS^{kn}h_n^B\ ,\ \ &&\kappa^\cH\equiv\kappa_{ijk}S^{il}h_l^BS^{jm}h_m^BS^{kn}h_n^B\ ,
\end{aligned}    
\end{equation}
where $S^{ij}$ is such that $S^{ij}S_{jk}=\delta^i_k$.


\section{Vacua from a quadratic superpotential}
\label{sec:Bilinear}

In this section, we present a generic classification of type IIB flux vacua at large complex structure arising from superpotentials that take a generic bilinear form, i.e., that are of the following kind:
\begin{equation}
W=\half\vec{Z}^t M \vec{Z} + \vec{L} \cdot \vec{Z} +Q\ ,
\label{eq:W_bilinear}
\end{equation}
where $\vec{Z}\equiv (\tau, \vec{z})$ and where the $(h^{2,1}+1)$-dimensional matrix $M$, the vector $\vec L$ and the scalar $Q$ are real flux-dependent quantities. Note that the matrix $M$ is symmetric by construction. 
As we will see in sect.~\ref{sec:IIB1_def}, the \emph{IIB1 scenario} that is of interest in this paper is precisely designed to get a quadratic structure from the superpotential \eqref{eq:Wfull}.
In the rest of the paper, we will apply the general formulas derived here in more detail and push the analytical developments. Note that generically the superpotential is cubic in the complex structure/axio-dilaton sector, as shown in the previous section.

Let us denote the covariant derivatives with respect to $\tau$ and $z^i$ in a vector notation $\vec D\equiv (D_\tau,D_i)$. Likewise, we package the first derivatives of the Kähler potential within the vector $\vec\partial K\equiv(K_\tau,K_i)$, which is pure imaginary and axion-independent (see eqs.~\eqref{eq:Kt} and \eqref{eq:Ki}). The vacuum equations then take the form
\begin{equation}
\vec D W=0\ \Longleftrightarrow\  M\vec Z +\vec L+ (\vec\partial K) W=0\ .
\label{eq:vacuum_eqs_bilinear}
\end{equation}
The superpotential at vacua enjoys a reality property. Indeed, decomposing $\vec Z\equiv B+i\vec T$ into eq.~\eqref{eq:W_bilinear} yields
\begin{equation}
\Im(W)=\vec B^t M\vec T+\vec L\cdot\vec T\ .
\end{equation}
On the other hand, and thanks to this expression for $\Im(W)$, the real part of \eqref{eq:vacuum_eqs_bilinear} contracted with $\vec T$ gives
\begin{equation}
\Im(W)\left(1+i\vec T\cdot\vec\partial K\right)=-\frac{4+\xi}{2(1+\xi)}\Im(W)=0\ .
\end{equation}
Here, we made use of eq.~\eqref{eq:Ki} and the definition of the LCS parameter $\xi$ introduced in eq.~\eqref{eq:def_xi} to express $\vec T\cdot\vec\partial K$. Since $\xi$ cannot be equal to $-4$, as explained below \eqref{eq:def_xi}, we deduce that $\Im(W)$ vanishes at vacua so that the superpotential is real on-shell.
With this result at hand, the vacuum equations \eqref{eq:vacuum_eqs_bilinear} split into
\begin{align}
M\vec B&=-\vec L\ ,\label{eq:B_gen}\\
M\vec T&=i(\vec\partial K) W\ ,\label{eq:T_gen}
\end{align}
which in particular imply that $\vec{L}$ should be in the image of the matrix $M$ in order to find a vacuum solution, which is a non-trivial requirement on the flux quanta when $M$ is not invertible. For this reason it is natural to discuss separately those cases in which the matrix $M$ is regular and when it is not. In both cases, 
using \eqref{eq:B_gen}, we can write the superpotential at vacua like
\begin{equation}
\label{eq:Wvac_int}
W=-\half\vec T^tM\vec T+Q'\ ,    
\end{equation}
where $Q'$ is a flux-dependent quantity defined by
\begin{equation}
\label{eq:Q'}
Q'\equiv Q - \half\vec L^t M^+ \vec{L}\ ,
\end{equation}
and $M^+$ is the generalized inverse of $M$, whose explicit expression we give below. Then, from \eqref{eq:T_gen} and \eqref{eq:Wvac_int} we deduce that
\begin{equation}
\label{eq:Wvac}
W=\frac{Q'}{1-\frac{i}{2}\vec T\cdot\vec\partial K}=\frac{4}{3}\frac{1+\xi}{\xi}Q'=-\frac{2}{3}e^{-K_{\rm cs}}\frac{Q'}{\Im\kappa_0}\ .    
\end{equation}
Therefore, when approaching the LCS point at $\xi=0$, the superpotential diverges. Also, notice that supersymmetric vacua are only possible if $Q'=0$.

\subsection[When $M$ is invertible]{\bm When $M$ is invertible}
\label{sec:Minvertible}
 When  $M$ has an inverse then $M^+ = M^{-1}$, and so eq.~\eqref{eq:B_gen} stabilizes all the axions at 
 \begin{equation}
 \vec B=-M^{-1}\vec L \ .\label{eq:B_nonsusy}
 \end{equation}
 On the other hand, eq.~\eqref{eq:T_gen} is implicit on the saxions since $\vec\partial K$ and $W$ depend on $\vec T$. This is summed up in the following expression for $\vec Z$:
\begin{equation}
\vec Z=-M^{-1}\left(\vec L+(\vec\partial K) W\right)\ .
\label{eq:Z}
\end{equation}
 The superpotential at vacua reads as \eqref{eq:Wvac} with  $Q'$ given by
\begin{equation}
Q'=Q-\half\vec L^tM^{-1}\vec L\ .
\end{equation}
As noted aobove, supersymmetric vacua only arise if $Q'=0$. But with $M$ invertible this would imply that $\vec T=\vec 0$ due to \eqref{eq:T_gen}. Supersymmetric vacua are thus forbidden when $M$ is regular.

\subsection[When $M$ is singular]{\bm When $M$ is singular}
\label{sec:Mnoninvertible}

As mentioned earlier, eq.~\eqref{eq:B_gen} tells us that $\vec L$ lies in the image of $M$ since $\vec{L}=M(-\vec B)$. As a consequence, the field directions inside the kernel of $M$ do not enter the superpotential. Thus, in the LCS approximation, the axionic directions that correspond to $\ker(M)$ do not enter the scalar potential at all, implying a number of flat directions. To describe the number of these flat directions one must distinguish between supersymmetric and non-supersymmmetric vacua:
    \begin{itemize}
        \item When $W\neq 0$, which corresponds to flux choices such that $Q' \neq 0$, we have that $\rank(M)$ of the axions are stabilized, while $h^{2,1}+1-\rank(M)$ constraints on the flux quanta must be satisfied in order for vacua to exist.
        To see this, we can diagonalize the matrix $M$ to a matrix $D\equiv\text{diag}(\lambda_0,\dots,\lambda_{r-1},0,\dots,0)$ with $\lambda_0,\dots,\lambda_{r-1}$ representing the $r\equiv\rank(M)$ non-zero eigenvalues of the matrix, and where there are as many zeroes as the dimension of the kernel. We write the similarity transformation with a matrix $N$ like
        \begin{equation}
        M= N^tDN\quad\text{ and }\quad N^t=N^{-1}\ .
        \end{equation}
        Defining $\vec B'\equiv N\vec B$ and $\vec L'\equiv N\vec L$, the axionic system of equations \eqref{eq:B_gen} becomes
        \begin{equation}
        D\vec B'=-\vec L'\ .
        \end{equation}
        We now split the $h^{2,1}+1$ indices $\{0,i\}$ like $\alpha\in\{0,\dots,r-1\}$ and $\beta\in\{r,\dots,h^{2,1}\}$ to get the following vacuum expectation values and constraints:
        \begin{equation}
        b'^\alpha=-\frac{\vec L'^\alpha}{\lambda_\alpha}\quad\text{ and }\quad \vec L'^\beta=0\ .
        \end{equation}

        The superpotential at vacua \eqref{eq:Wvac} involves the quantity $Q'$ which again is flux-dependent-only and reads
        \begin{equation}
        \label{eq:Q'sing}
        Q'=\half \vec L' \cdot\vec B '+Q=-\half \sum_\alpha\frac{(\vec L'^\alpha)^2}{\lambda_\alpha}+Q = - \half \vec{L}^t M^+ \vec{L}+ Q \ ,
        \end{equation}
        where $M^+= N^t D^+ N$ and $D^+ \equiv\text{diag}(\lambda_0^{-1},\dots,\lambda_{r-1}^{-1},0,\dots,0)$.
        As for the saxions, they satisfy the non-linear implicit relation \eqref{eq:T_gen}, where the superpotential $W$ takes the saxion-dependent form \eqref{eq:Wvac}. Since all axions enter in this condition, one generically expects that its solution stabilizes all of them. 
        \item When $W=0$, we read from \eqref{eq:vacuum_eqs_bilinear} that the vacuum solutions are
        \begin{equation}
        \label{eq:sol_susy}
        \vec Z=\vec B+\ker\, (M)\ ,
        \end{equation}
        and so only $\rank(M)$ complex moduli are stabilized. As in the previous case, the same $h^{2,1}+1-\rank(M)$ constraints on the flux quanta should hold. Moreover, $Q'=0$ provides one additional constraint on the fluxes. In total, we expect the fluxes to satisfy $h^{2,1}+2-\rank (M)$ relations in order to fall into this supersymmetric class of vacua.
    \end{itemize}

\subsection{The IIB1 family}
\label{sec:IIB1_def}

In this subsection, we introduce the \emph{IIB1 scenario} described in \cite{Marchesano:2021gyv}. There, the starting point of the authors is F-theory compactifications at large complex structure. They develop analytical expressions of the scalar potential in full generality and recast it with a bilinear structure $V=\rho_AZ^{AB}\rho_B$, which is found to be very useful to express the vacuum equations and systematically characterize the possible families of vacua (see \cite{Bilinear1,Bilinear2,Bilinear3,Bilinear4,Bilinear5,Bilinear6,Bilinear7} for applications of this strategy). Requiring the tadpole not to diverge at LCS, two distinct families of vacua are uncovered and, in the type IIB limit of F-theory, they yield two scenarios, one of them being the \emph{IIB1} setup  on which we focus here. It is characterized by putting some specific flux quanta to zero:
\begin{equation}
\label{eq:def_IIB1}
\text{IIB1 flux configuration:}\quad f_A^0=0\ ,\  h_A^0=0\ \text{ and }\ h_A^i=0\ ,\ i\in\{1,\dots,h^{2,1}\}\ . 
\end{equation}

We can motivate the interest on this ansatz by looking at its effects on the type IIB superpotential \eqref{eq:Wfull}. The choice $f_A^0 = h_A^0 = 0$, i.e. $N_A^0 = 0$, has important consequences. We see that it removes the ``pure complex structure'' cubic, highest-order term $z^iz^jz^k$, from the superpotential. This ends up being quite a non-trivial effect, since it leads to solutions arbitrarily close to the LCS point, as opposed to the $N_A^0 \neq 0$ case \cite{Sousa:2014qza,Marsh:2015zoa,Brodie:2015kza}. In one-parameter models, this choice of fluxes has been proven to lead to completely different mass spectra than in the generic $N_A^0 \neq 0$ case, along with its own statistical ensembles of vacua \cite{Blanco-Pillado:2020wjn}. Following a similar reasoning as to the statements above, we remark that with the additional choice $h_A^i=0$ we get $N_A^i = f_A^i$, which removes the mixed (complex structure and axio-dilaton) cubic term $z^i z^j \tau$ from the superpotential, and only leaves a quadratic one on $z^i z^j $. 

Thus, the IIB1 flux choice ensures that the superpotential takes the bilinear form \eqref{eq:W_bilinear} with $\vec{Z}^t = (\tau, \vec{z}^t)$ and the following flux-dependent quantities:
\begin{equation}
\label{eq:MLQ}
M\equiv\begin{pmatrix}
0 & -\vec h^{B\, t}\\
-\vec h^{B} & S_{ij}
\end{pmatrix},
\quad \vec L\equiv(-h_0^B,f_i^B+a_{ij}f_A^j)\ ,\quad Q\equiv f_0^B-c_if_A^i\ ,
\end{equation}
and where the matrix $S$ is defined as $S_{ij}\equiv\kappa_{ijk}f_A^k$. We further write $\vec L\equiv (L_0,L_i)$ so that $L_0\equiv -h_0^B$ and $L_i\equiv f_i^B+a_{ij}f_A^j$. Note that in the following sections, we will focus on flux configurations for which the matrix $S$ is invertible. When it is the case, the invertibility of $M$ is determined by the value of $\det(M)/\det(S)\equiv\cH= h_i^BS^{ij}h_j^B$.

In \cite{Marchesano:2021gyv}, the authors expressed the vacuum equations descending from the F-theory ones and wrote them at first order in the LCS parameter $\xi$.
In the following, we will generalize this analysis and extend it to the full LCS region, i.e. for arbitrary $\xi$, by applying the generic results of the present section. We consider the non-supersymmetric (sect.~\ref{sec:nonsusy}) and supersymmetric (sect.~\ref{sec:susy}) vacua highlighted above and, in both cases, fully analytical relations for the axions and saxions vacuum locations are displayed. In the non-supersymmetric case, the analytical control over the saxions comes at the cost of restricting to a particular branch of solutions that we know is not unique thanks to numerics. Moreover, yet in a further subclass, we are able to express the vacuum expectation values with formulas that are exact in $\xi$ and we are able to uncover the scalar mass spectrum analytically.


\section{Non-supersymmetric vacua}
\label{sec:nonsusy}

We study here the non-supersymmetric flux vacua exhibited in the previous section, that can arise both with $M$ invertible or singular. We recall that the vacuum equations reduce to \eqref{eq:B_gen} and \eqref{eq:T_gen}, where the superpotential at vacua takes the form \eqref{eq:Wvac}. We thus have
\begin{align}
M\vec B&=-\vec L\ ,\label{eq:B_nonsusybis}\\
M\vec T&=-\frac{2}{3}ie^{-K_{\rm cs}}\frac{Q'}{\Im\kappa_0}(\vec\partial K)\ .\label{eq:T_nonsusy}
\end{align}
We first focus on the saxionic system which can be recast as
\bea
\label{dilaton}
- 3 h^B_i t^i t^0 &=& e^{-K_{\rm cs}} \frac{Q'}{\Im \kappa_0} =  \frac{4}{3} \frac{Q'}{\Im \kappa_0} \kappa_{ijk}t^it^kt^k -2Q'\ ,\label{eq:Q/k}\\
- h^B_i t^0 + S_{ij} t^j  &=&  \frac{4}{3} \frac{Q'}{\Im \kappa_0} \kappa_{ijk}t^jt^k\ ,
\label{saxions}
\eea
from which it seems natural to define the following rescaled variables:
\be
x^0 \equiv  \frac{4}{3} \frac{Q'}{\Im \kappa_0} t^0\ , \qquad x^i \equiv  \frac{4}{3} \frac{Q'}{\Im \kappa_0} t^i \ . 
\ee

In terms of these rescaled variables, the above equations read
\bea
\label{dilatonx}
- 3 h^B_i x^i x^0 &=&   \kappa_{ijk}x^ix^kx^k - \S\alpha\ ,\\
- h^B_i x^0 + S_{ij} x^j  & = &\kappa_{ijk}x^jx^k\ ,
\label{saxionsx}
\eea
where
\be
\label{eq:def_alpha}
\alpha \equiv \frac{2^5 Q'^3 }{ 3^2 (\Im \kappa_0)^2\S}\ , \qquad \S \equiv \kappa_{ijk} f_A^if_A^jf_A^k\ .
\ee
Notice that eq.~\eqref{saxionsx} only depends on triple intersection numbers and fluxes bounded by the D3-brane tadpole. Therefore, one expects $x^A \sim {\cal O}(N_{\rm flux}^{1/2})$, with $A\in\{0,i\}$ and $\Nf=-f_A^ih_i^B$. To generate larger values for the saxions $t^A$, one may consider flux choices such that
\be
\frac{Q'}{\Im \kappa_0} \ll 1\ .
\ee
When it is the case,
\be
1 \gg |\alpha| \simeq |\xi|\ ,
\ee
so we are in a large complex structure regime. 

The system of equations \eqref{dilatonx} and \eqref{saxionsx} is rather involved as it is, so we will propose an ansatz to make analytical progress, that we will further refine into two cases in which we are able to obtain concrete results. Our working assumption will be that the matrix $S$ is invertible, and we will oftentimes also assume that $\S \neq 0$, in order to define $\alpha$ as above. To build the ansatz we take inspiration from the analysis performed in \cite{Marchesano:2021gyv}. There, a decomposition of the flux quanta $f^i_A$ and $h_i^B$ in terms of saxion vevs was introduced  as follows
\begin{equation}
    f^i_A = A t^i + C^i\ , \qquad h_i^B = B \kappa_{ijk}t^jt^k + C_i\ ,
    \label{decomp}
\end{equation}
with $C^i \kappa_{ijk}t^jt^k = C_it^i =0$. This fully general decomposition was helpful in the study of the equations of motion, which required the relations $A=t^0 B$ and $-C_i t^0=\kappa_{ij}C^j$. However, in order to provide concrete expressions for the vacuum expectation values of the moduli including first order polynomial corrections, the authors restricted the flux space to the case $C_i=C^i=0$ and linearized the equations in $\xi$. We now aim to extend this ansatz and to consider the effect of polynomial corrections at all orders. To do so we turn on the vector $C^i$ but demand a concrete relation with the flux quanta. We thus propose the ansatz
\be
\label{eq:ansatz}
t^i \equiv \hat{t} f_A^i + \tilde{t} S^{ij} h_i^B  \quad\implies\quad x^i \equiv \hat{x} f_A^i + \tilde{x} S^{ij} h_i^B\ .
\ee
The vacua equations then read
\bea
\label{dilaton2}
3 x^0 \left( N_{\rm flux} \hat{x} - {\cH}\tilde{x}\right) & = & \hat{x}^3 \S - 3N_{\rm flux} \hat{x}^2 \tilde{t} + 3{\cH} \hat{x} \tilde{x}^2+ \tilde{x}^3  \kappa^{\cH} - \S \a\ ,\\
h^B_i (\tilde{x} -x^0) + S_i \hat{x}   & = & \hat{x}^2 S_i + 2\hat{x}\tilde{x} h_i^B  + \tilde{x}^2 \kappa^{\cH}_i\ ,
\label{saxions2}
\eea
where we have defined
\be
\kappa^{\cH}_i \equiv \kappa_{ijk}S^{jl}S^{km}   h_l^B h_m^B\ , \qquad \kappa^{\cH} \equiv \kappa_{ijk} S^{il} S^{jm}S^{kn}   h_l^B h_m^B  h_n^B\ ,
\ee
and recall that $\cH\equiv\det(M)/\det(S)=h_i^BS^{ij}h_j^B$. Upon contracting \eqref{saxions2} with $f_A^i$ and with $S^{ij}h_j^B$, and plugging back into \eqref{saxions2}, we obtain a consistency flux condition that reads
\be
\left(N_{\rm flux}^2 - \S {\cH} \right)  \kappa^{\cH}_i  + \left(\S \kappa^{\cH}  + {\cH} N_{\rm flux} \right) h^B_i + \left( \kappa^{\cH} N_{\rm flux}  +  {\cH}^2 \right)S_i =0\ ,
\label{flux1}
\ee
where $S_i\equiv \kappa_{ijk}f_A^jf_A^k$.

As evoked above, progressing without further refining the branch under consideration seems very involved. However, we notice that the constraint \eqref{flux1} is compatible with the relation $\cH=0$, which will define our first subclass of interest developed in sect.~\ref{sec:S0}. This case falls into the kind of non-supersymmetric vacua described in sect.~\ref{sec:Mnoninvertible} where the matrix $M$ is singular. The other subclass to be studied in the sequel assumes the ansatz \eqref{eq:ansatz} with the simplification $\tilde t=0$, and will be discussed in sect.~\ref{sec:simpler}

\subsection[A subcase with $M$ singular]{\bm A subcase with $M$ singular}
\label{sec:S0}

In this subsection, we push the analytics sketched above with the further flux condition
\begin{equation}
\cH=h_i^BS^{ij}h_j^B=0\ .    
\end{equation}
In this case, the matrix $M$ has a one-dimensional kernel generated by $\langle (1,S^{ij}h_j^B)\rangle$. From the generic discussion of sect.~\ref{sec:Mnoninvertible}, we then expect one constraint to arise from the axionic system \eqref{eq:B_nonsusybis} as well as one flat direction. More precisely, we have
\begin{align}
h_i^BS^{ij}L_j=h_0^B\quad\text{ and }\quad b^i=-S^{ij}L_j+b^0S^{ij}h_j^B\ .
\end{align}

The saxionic system given by eqs.~\eqref{dilaton2} and \eqref{saxions2} reduces to the following one when $\cH=0$:
\bea
3 N_{\rm flux} \hat{x} x^0& = & \hat{x}^3 \S - 3N_{\rm flux} \hat{x}^2 \tilde{t}+ \tilde{x}^3  \kappa^{\cH} - \S \a\ ,\\
h^B_i (\tilde{x} -x^0) + S_i \hat{x}   & = & \hat{x}^2 S_i + 2\hat{x}\tilde{x} h_i^B  + \tilde{x}^2 \kappa^{\cH}_i\ ,
\eea
and the flux condition \eqref{flux1} becomes\footnote{Notice that this condition is automatically satisfied for models with two complex structure moduli where $\cH =0$, because then the vector in \eqref{fluxcond} is always orthogonal to $f_A^i$ and $S^{ij}h^B_j$. }
\be
N_{\rm flux}^2 \kappa^{\cH}_i  + \S \kappa^{\cH}h^B_i + \kappa^{\cH} N_{\rm flux}S_i =0\ .
\label{fluxcond}
\ee
One can manipulate the system of equations to arrive at an expression giving $\tilde x$ as a function of $\hat x$, a relation giving $x^0$ as a function of $\hat x$ and $\tilde x$ and an equation involving only $\hat x$. Indeed we have\footnote{These expressions assume $\kappa^\cH\neq0$ and $\S\neq 0$. If not, we find $\hat x=1$, $\tilde x=-x^0$ and one saxion is left unstabilized. When $\kappa^\cH=0$ and $\S\neq 0$, the flux relation $\alpha=1$ should also be satisfied.}
\begin{align}
&\tilde x^2=\frac{\Nf}{\kappa^\cH}\hat x(\hat x-1)\ ,\label{eq:xtilde}\\
&x^0=\frac{\S\hat x(\hat x-1)}{\Nf}+\tilde x-2\hat x\tilde x\ ,\label{eq:x0}\\
&\left(2\hat x^3-3\hat x^2+\alpha\right)^2=16\frac{\Nf^3}{\S^2\kappa^\cH}\hat x^3(\hat x-1)^3\ .\label{eq:sixth}
\end{align}

The last equation involving only $\hat x$ is polynomial of sixth order. To proceed, we can neglect $\alpha$ to find approximate solutions valid close to the LCS point. The polynomial then becomes only of third order and can be written like
\be
 \hat{x}^3 - 3  \hat{x}^2  + 3\frac{\beta - 3/4}{\beta-1}  \hat{x} - \frac{\beta}{\beta-1} \simeq 0\ , \quad\text{ with }\quad \beta \equiv 4 \frac{N_{\rm flux}^3}{\S^2\kappa^{\cH}}\ .
\ee 
This cubic equation admits three roots, either one real and two complex or three reals. If we label them $\hat x_0$, $\hat x_1$ and $\hat x_2$, they are given by
\begin{equation}
\hat x_k=1 + \frac{j^k\gamma}{2} -\frac{1}{2j^k\gamma(\beta-1)}\ ,\quad k\in\{0,1,2\}\quad \text{ and }\quad j\equiv \frac{-1+i\sqrt{3}}{2}\ ,\label{eq:xk}
\end{equation}
and where $\gamma$ is such that
\begin{equation}
\gamma^3\equiv \frac{1}{\beta-1}\left(1+\sqrt{\frac{\beta}{\beta-1}}\right)\ .
\end{equation}
Note that we cannot determine in full generality which of these solutions correspond to the real ones. With a solution for $\hat x$, eq.~\eqref{eq:xtilde} allows to compute $\tilde x$ so that we can deduce $x^i$ from the ansatz. On the other hand, eq.~\eqref{eq:x0} allows to compute $x^0$. From the definitions of the rescaled variables, one can then deduce the vacuum expectation values of the saxions $t^0$ and $t^i$.

We can refine this approximate solution, valid near the LCS point, by using a pertubative approach. Indeed, if we denote the above approximate solution $\hat x^{(0)}$, we can write
\begin{equation}
\hat x=\hat x^{(0)}+\delta\hat x\ ,\label{eq:xkrefined}
\end{equation}
with $\delta\hat x\sim\mathcal{O}(\alpha)\ll 1$. Plugging this into the full equation \eqref{eq:sixth} and restricting to first order in $\alpha$ yields
\begin{equation}
\delta\hat x=\frac{(2\hat x^{(0)}-3)\S^2\kappa^\cH \alpha }{6(\hat x^{(0)}-1)\left[4\Nf^3(\hat x^{(0)}-1)(2\hat x^{(0)}-1)-\S^2\kappa^\cH\hat x^{(0)}(2\hat x^{(0)}-3)\right]} + {\cal O}(\alpha^2)\ . 
\end{equation}
One can plug this refined value of $\hat{x}$ into \eqref{eq:sixth}, and again linearize the equation to obtain its value to the next order in $\alpha$. The procedure can be repeated to provide an analytic expression up to any order in $\alpha$.

\subsection{A simpler ansatz for full analyticity}
\label{sec:simpler}

Another very interesting subclass of vacua arises when one considers a particular restriction of the ansatz proposed in \eqref{eq:ansatz}. This restriction consists in assuming $\tilde t=0$, so that we are left with
\begin{equation}
\label{eq:that}
t^i\equiv \hat t f_A^i\quad\implies\quad x^i \equiv \hat{x} f_A^i\ .
\end{equation}
For reasons that will be clearer later, we call this branch of vacua the \emph{no-scale aligned} branch. The vacuum equations for this branch reduce to
\bea
\label{dilaton1}
3 N_{\rm flux} \hat{x} x^0 &=&   \hat{x}^3 \S - \S \a\ ,\\
- h^B_i x^0 + S_{i} \hat{x}  & = & S_i \hat{x}^2\ .
\label{saxions1}
\eea
Contracting \eqref{saxions1} with $f_A^i$ we obtain 
\be
\S (\hat{x}^2 -\hat{x}) = N_{\rm flux} x^0\ ,
\label{saxions1c}
\ee
so we deduce that $\S \neq 0$. Plugging this equation back into \eqref{saxions1}, we obtain a condition for the flux vector $\vec{h}_B$:
\be
 h^B_i = - N_{\rm flux}\frac{S_i}{\S} \quad \Longrightarrow \quad h_i^B = -  \hat{h}^B \frac{S_i}{q}\ ,
\label{hBicond}
\ee
where $\hat{h}^B \in \mathbb{Z}$ and $q \equiv \gcd (S_i)$. This flux relation can be thought of as a simpler version of \eqref{flux1} for this particular ansatz. It is worth noting that in the language of \cite{Marchesano:2021gyv},  \eqref{eq:that} and \eqref{hBicond} correspond to the choice $C_i$, $C^i=0$ using the decomposition \eqref{decomp}. If we  assume that the matrix $S$ is invertible, the above relation implies that $\cH \equiv h^B_iS^{ij} h^B_j \neq 0$, and so $M$ is regular. We are thus in the generic case described in sect.~\ref{sec:Minvertible}. In the sequel, we will solve the axionic and saxionic systems of equations.

\subsubsection{Moduli stabilization}

\paragraph{Axions:}

The axions are stabilized at $\vec B=-M^{-1}\vec L$. The inverse of the matrix $M$ defined in eq.~\eqref{eq:MLQ} cannot be expressed in full generality but it can under the assumption that the matrix $S$ is invertible.\footnote{And in this case we saw above that $\cH\neq0$.} When it is the case, we have \cite{Marchesano:2021gyv}
\begin{equation}
\label{eq:Minv}
M^{-1}=\frac{1}{\cH}\begin{pmatrix}
-1 & -S^{jk}h_k^B\\
-S^{ik}h_k^B & \cH S^{ij}-S^{ik}S^{jl}h_k^Bh_l^B
\end{pmatrix}.
\end{equation}
This yields
\begin{align}
\begin{split}
\label{eq:bs}
b^0 &= \frac{h_i^B S^{ij} L_j - h_0^B}{\cH}\ ,\\
b^i &= S^{ij} \left( b^0 h_j^B - L_j \right) \ .
\end{split}
\end{align}
Note that the quantity $Q'$ in this case is given by
\begin{align}
	Q' = f_0^B - f_A^i c_i + \frac{(h_i^B S^{ij} L_j - h_0^B)^2}{2 h_i^B S^{ij} h_j^B} - \frac{1}{2} L_i S^{ij} L_j\ .
	\label{eq:rho}
\end{align}

\paragraph{Saxions:}

For the saxions, the relation \eqref{saxions1c} allows to solve for $\hat x$ as a function of $x^0$. We find
\be
\hat{x} = \frac{1}{2} \left( 1 \pm \sqrt{1 + 4 \frac{N_{\rm flux}}{\S} x^0}  \right)\ .
\label{premaster}
\ee
We now plug \eqref{dilaton1} into this expression, to obtain
\be
2\hat{x} = 1 \pm \sqrt{ 1 + \frac{4}{3}\left(\hat{x}^2-\frac{\a}{\hat{x}}\right)}\ ,
\ee
which yields the following cubic equation:
\be
2\hat{x}^3 -3\hat{x}^2 + \a = 0\ .
\label{cubic0}
\ee

The discriminant $\Delta$ of the cubic can be expressed simply as a function of $\alpha$ like
\begin{equation}
\Delta=4\alpha(\alpha-1)\ .
\end{equation}
When $\alpha<0$ or $\alpha>1$, the discrimant is positive and there is a single real root given by
\begin{equation}
\label{sol}
\hat x=\half\left(1+\Gamma+\frac{1}{\Gamma}\right)\quad\text{ where }\quad \Gamma^3\equiv 1-2\left(\alpha+\sqrt{\alpha(\alpha-1)}\right)\ . 
\end{equation}
When $\alpha\in[0,1]$, the discriminant is negative and there are three real roots. The formula above is still valid to describe one of them if one defines the square and cubic roots as principal values. A (unique or not) solution for $\hat t$ is thus always given by
\begin{equation}
\label{eq:solthat}
\hat t=\frac{3}{8}\frac{\Im\kappa_0}{Q'}\left(1+\Gamma+\frac{1}{\Gamma}\right)\ .
\end{equation}
We will show below that this expression for $\hat t$ with roots defined as principal values always gives the unique physical solution. With this exact expression for $\hat t$ at hand, we can use \eqref{premaster} to isolate $t^0$. With the help of eq.~\eqref{eq:Q/k} that we repeat here
\begin{equation}
\label{eq:Q/kbis}
3\Nf\hat tt^0=e^{-K_{\rm cs}}\frac{Q'}{\Im\kappa_0}\ ,
\end{equation}
we arrive at
\begin{align}
\label{eq:t0t}
    t^0 = 
    \frac{q}{\hat h^B}
    \frac{2 \S \hat t^3-3\Im\kappa_0}{4\S\hat t^3+3\Im\kappa_0} \ \hat{t}\ .
\end{align}
Using \eqref{eq:solthat}, we can express a useful relation between the LCS paramater $\xi$ and the quantity $\alpha$:
\begin{align}
    \label{eq:flux_const_away}
    &\frac{\xi}{(\xi-2)^3}=\frac{\alpha}{27}\ .
\end{align}

\paragraph{Physical solutions:}

Let us now take a more detailed look at the physical solutions depending on the sign of $\alpha$. From eq.~\eqref{eq:Q/kbis} above, we see that the sign of $\hat t$ is the same as that of the ratio $Q'/\Im\kappa_0$. We thus have:
\begin{itemize}
    \item When $\alpha<0$, then if $\hat t>0$ we deduce $Q'<0$ from the definition of $\alpha$ and thus $\Im\kappa_0<0$ from \eqref{eq:Q/kbis}. If $\hat t<0$ we deduce $Q'>0$ from the definition of $\alpha$ and still $\Im\kappa_0<0$ from \eqref{eq:Q/kbis}. Thus, $\alpha$ negative corresponds exclusively to models with a negative $\Im\kappa_0$. For those models, we mentioned in sect.~\ref{sec:Notations} that $\xi$ should be in the range $[0,1/2]$ for the Kähler metric to be well-defined with positive eigenvalues. By solving $\xi<1/2$, we can deduce a lower bound that the solution $\hat x$ of the cubic should satisfy. We find
    \begin{equation}
    \hat x>2^{1/3}|\alpha|^{1/3}\ .
    \end{equation}
    Equivalently, \eqref{eq:flux_const_away} yields $\alpha>-4$.
    \item When $\alpha>0$, same arguments lead to conclude that no matter what the sign of $\hat t$ is, $Q'$ has the same and $\Im\kappa_0$ is positive. For those models, we should have $\xi\in[-1,0]$. Solving $\xi>-1$, we find
    \begin{equation}
    \hat x>|\alpha|^{1/3}\ ,
    \end{equation}
    and equivalently, \eqref{eq:flux_const_away} yields $\alpha<-1$.
\end{itemize}

\begin{figure}[!h]
\centering
\includegraphics[scale=0.45]{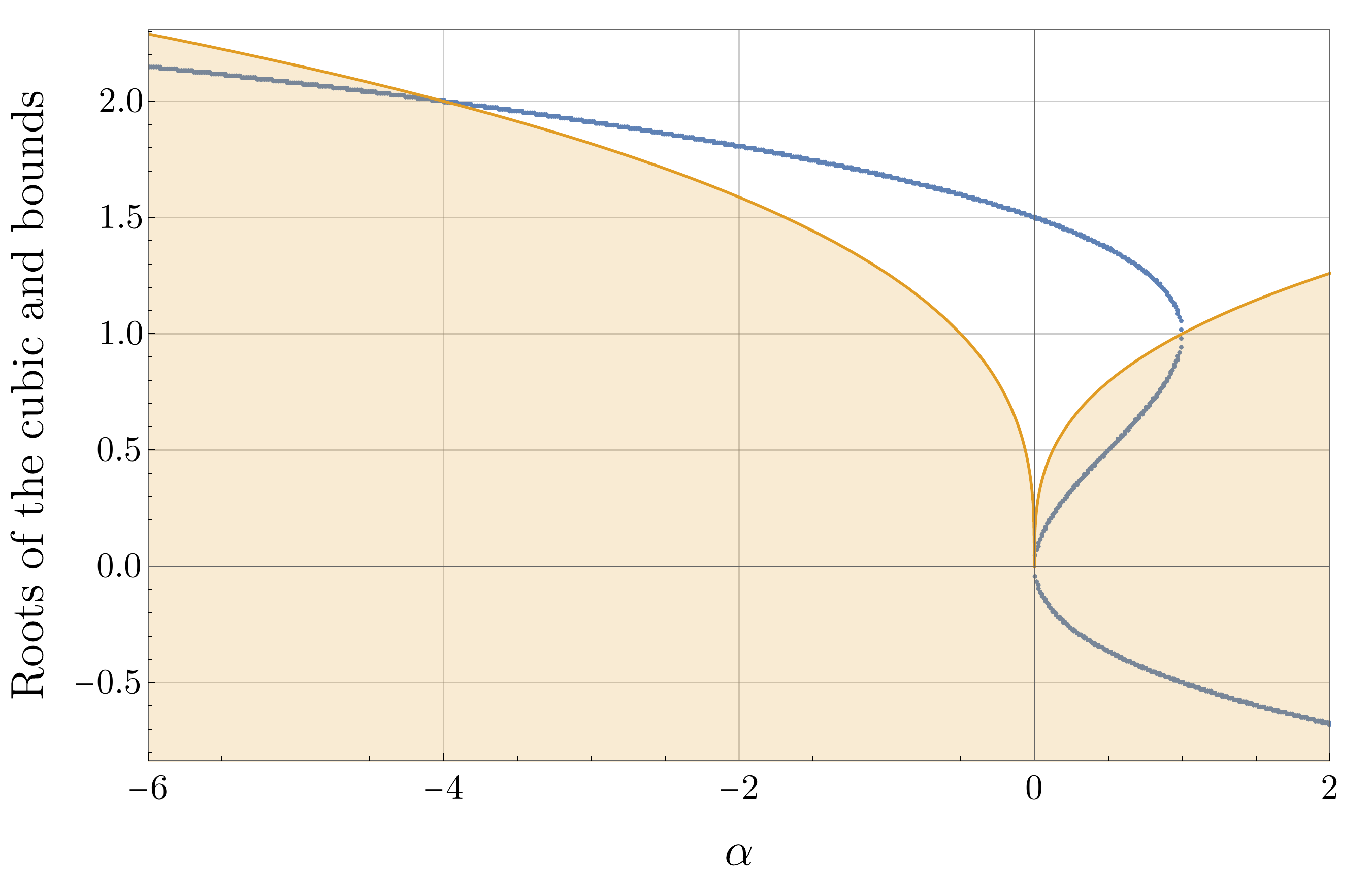}
\caption{The roots of the cubic \eqref{cubic0} with respect to the parameter $\alpha$.}
\label{roots}
\end{figure}

Figure \ref{roots} shows the values of the roots of the cubic equation \eqref{cubic0} as a function of $\alpha$ as well as the bounds derived above. We observe that for $\alpha<0$, the Kähler cone bound is violated when $\alpha<-4$ and when $\alpha>1$, there is no physical solution as expected. When $0<\alpha<1$, we observe that only one root is compatible with the Kähler cone condition. Moreover, it turns out that this is the one that can be expressed like \eqref{sol} with the proper principal value definitions of the roots.

Apart from the full analytical expressions for the moduli vacuum expectation values, valid at arbitrary $\xi$, the simpler ansatz under consideration here also allows to uncover the scalar mass spectrum. Computing these masses is the purpose of the next subsection.

\subsubsection{Mass spectrum}
\label{sec:masses}

To uncover the mass spectrum, we make use of the symplectic decomposition of the flux vector introduced in \cite{Denef:2004ze}, which reads
\begin{equation}
    N = \sqrt{4\pi} e^{K_{\rm cs}} \left( -i W \bar{\Pi} + 2 t^0 D_{\bar{\tau}} D_{\bar{j}} \bar{W} K^{\bar{j}i} D_i \Pi \right)\ .
    \label{eq:N_structure}
\end{equation}
Inserting the flux constraints of the IIB1 setup $f_A^0=h_A^0=h_A^i=0$ inside the above expression yields two relations
\begin{equation}
W = -2i t^0 D_{\bar{\tau} \bar{j}} \bar{W} K^{\bar{j}i} K_i\ \quad\text{ and }\quad f_A^i= 2 e^{K_{\rm cs}} \left( t^0 D_{\bar{\tau} \bar{j}} \bar{W} K^{\bar{j}i} - t^i W \right)\ ,
\end{equation}
from which we deduce
\begin{equation}
\label{eq:DtW_Int}
D_{\tau i} W = \frac{1}{2t^0} K_{i\bar{j}} \left( e^{-K_{\rm cs}} f_A^j + 2 t^j \bar{W} \right)\ .
\end{equation}
Now we can make use of the proportionality relations \eqref{eq:that} that defines the ansatz to replace $f_A^j$ in the above formula and factor a term $K_{i\bar j}t^j$. From eqs.~\eqref{eq:Ki} and \eqref{eq:Kij}, this factor reads
\begin{align}
    K_{i\bar{j}} t^j = - 2 \mathring{\kappa}_{ijk} t^j t^k + 4 \mathring{\kappa}_{imn} \mathring{\kappa}_{jpq} t^m t^n t^j t^p t^q = i \left( 1 - 2 \mathring{\kappa} \right) K_i\ ,
\end{align}
where we have defined $\mathring{\kappa} \equiv e^{K_{\rm cs}} \kappa_{ijk} t^i t^j t^k$. Plugging this result back into eq.~\eqref{eq:DtW_Int} yields
\begin{equation}
D_{\tau i} W = \frac{ i \left( 1 - 2 \mathring{\kappa} \right)}{2t^0} \left( 2 \bar{W} -  \frac{e^{-K_{\rm cs}}}{\hat t}\right)K_i\ .
\end{equation}

\begin{figure}[!t]
\centering
\includegraphics[scale=0.45]{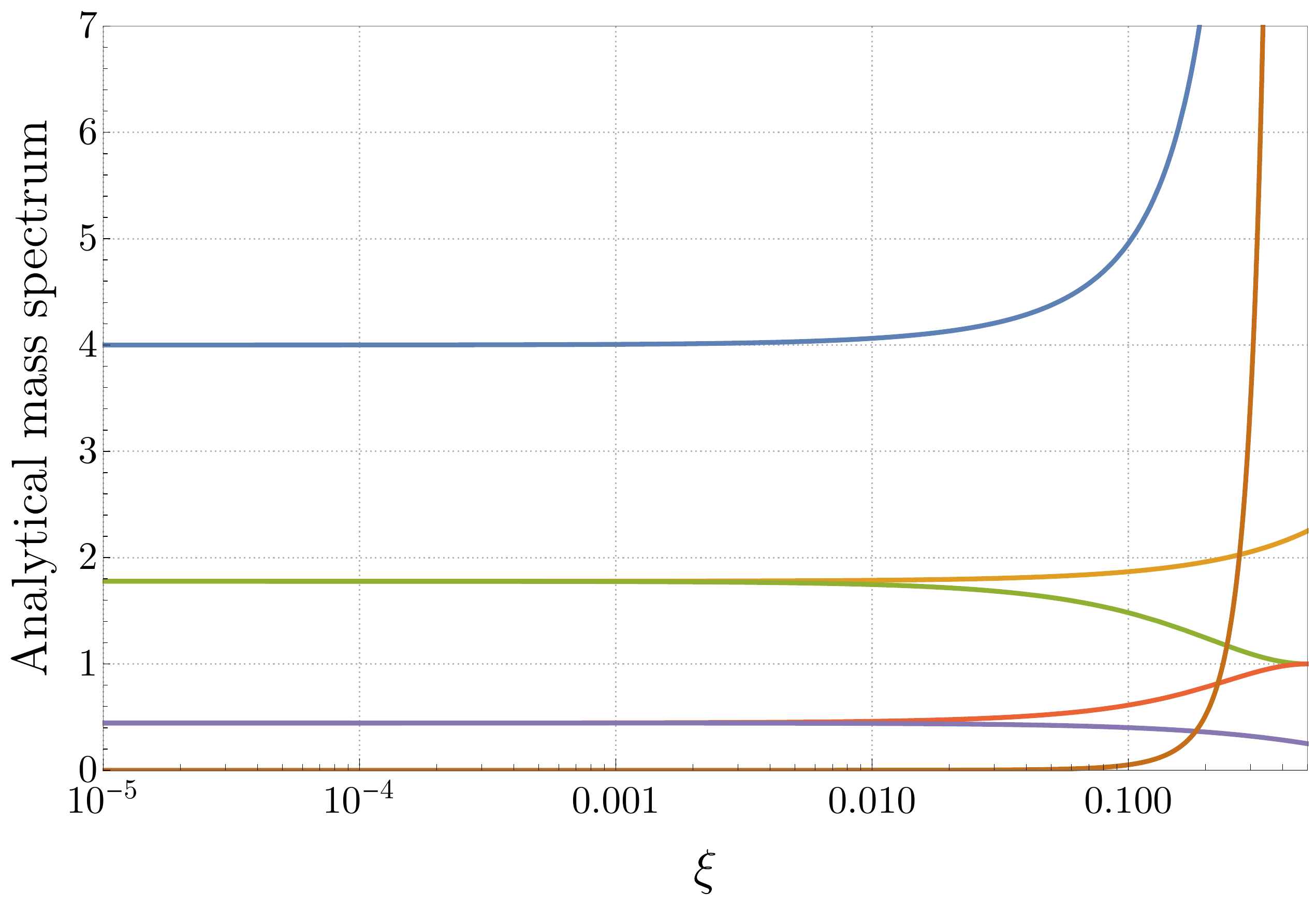}
\caption{Evolution of the scalar mass spectrum \eqref{eq:nsa_mass_spectrum} with respect to the LCS parameter $\xi$. The correspondence between the curves and the labels $\pm\lambda$ of the different modes is as follows: Blue curve is $+1$; Orange curve is $+\lambda$, $\lambda=2,\dots,h^{2,1}$; Green curve is $+0$; Red curve is $-0$; Purple curve is $-\lambda$, $\lambda=2,\dots,h^{2,1}$ and Brown curve is $-1$.}
\label{analytical_mass_spectrum}
\end{figure}

These steps show that under the IIB1 flux configuration and for our branch of solution of interest, the two-derivative of the superpotential with respect to the axio-dilaton and some complex structure field is proportional to the first derivative of the Kähler potential with respect to this latter modulus. As such, the IIB1 scenario fullfills the prerequisite for the derivation of the \emph{no-scale aligned} mass spectrum, introduced in \cite{Blanco-Pillado:2020hbw} and reviewed in appendix~\ref{sec:NSA}. The tree-level mass spectrum is thus given by \eqref{eq:nsa_mass_spectrum_app} that we repeat here:

\begin{equation}
	\frac{\mu^2_{\pm \lambda}}{m_{3/2}^2} = 
	\left\lbrace
	\begin{array}{ll}
		\left( 1 \pm \sqrt{\frac{1 -2 \xi}{3}} \hat{m} (\xi) \right)^2  & \lambda = 0  \\
		\left( 1 \pm \sqrt{\frac{1 -2 \xi}{3}} (\hat{m} (\xi))^{-1} \right)^2  & \lambda = 1 \\
		\left( 1 \pm \frac{1+\xi}{3} \right)^2 & \lambda = 2, \ldots, h^{2,1}
	\end{array}
	\right.
	\label{eq:nsa_mass_spectrum}
\end{equation}
where we have defined the quantities
\begin{align}
\begin{split}
&\hat{m} (\xi) \equiv \frac{1}{\sqrt{2}} \left( 2 + \kappa (\xi)^2 - \kappa (\xi) \sqrt{4 + \kappa (\xi)^2} \right)^{1/2}\ ,\\
&\kappa (\xi)\equiv 2 (1+ \xi)^2 / \sqrt{3(1-2\xi)^3}\ .
\end{split}    
\end{align}
The evolution of this normalized mass spectrum is displayed in fig.~\ref{analytical_mass_spectrum}.
Expanded around the LCS point at $\xi=0$, the spectrum reads
\begin{equation}
	\frac{\mu^2_{\pm \lambda}}{m_{3/2}^2} = 
	\left\lbrace
	\begin{array}{ll}
		\frac{16}{9}+\mathcal O(\xi)\ ,\  \frac{4}{9}+\mathcal O(\xi)& \lambda = 0  \\[5pt]
		4+\mathcal O(\xi)\ ,\ \frac{9}{4}\xi^2+\mathcal O(\xi^3)  & \lambda = 1 \\[5pt]
		\frac{16}{9}+\mathcal O(\xi)\ ,\  \frac{4}{9}+\mathcal O(\xi) & \lambda = 2, \ldots, h^{2,1}
	\end{array}
	\right.
\end{equation}
Notice that the mode labeled by $-1$ becomes rapidly massless as $\xi\to 0$, as can also be seen from fig.~\ref{analytical_mass_spectrum}. This is also true in Planck units, since the gravitino mass dependence on $\xi$ is given by
\begin{equation}
m_{3/2}^2=\frac{3}{2\V^2}\frac{\S\hat h^B}{q(2-\xi)} M_{\rm P}^2=\frac{3}{2\V^2}\frac{\Nf}{2-\xi}M_{\rm P}^2\ . 
\end{equation}
This nicely matches the expectations put forward in \cite{Marchesano:2021gyv}. There it was found that given the choice of fluxes \eqref{hBicond},  polynomial corrections are required to stabilize all moduli, and that otherwise a field is left unstabilized. It is thus natural to identify such a field with the lightest mode of the spectrum, whose mass goes proportional to $\xi$ as we approach the LCS point. 

All these results are verified by appendix \ref{ap:spot}, which develops a  different approach to the computation of the mass spectrum. This method works directly with the scalar potential derived from the results of \cite{Marchesano:2021gyv},  from where the Hessian can be obtained. One can see that in terms of the Hessian, the axion-like fields and their saxionic partners are decoupled. Therefore, by analyzing one of these two sets, it enables us to distinguish between axions and saxions in \eqref{eq:nsa_mass_spectrum}. In particular,  appendix \ref{ap:spot} works out explicit analytic expressions for the axionic masses of the no-scale aligned branch, obtaining a perfect match with half of the spectrum in \eqref{eq:nsa_mass_spectrum}. One can then check that the lightest field of \eqref{eq:nsa_mass_spectrum} is not one of the axion-like fields and that it instead belongs to the saxionic sector, in agreement with the expectations of \cite{Marchesano:2021gyv}.

\subsubsection{Generating flux vacua}
\label{sec:generate_vacua}

In the previous paragraphs we have studied how a choice of fluxes which satisfies
\begin{align}
	f_A^0 = h_A^0 = h_A^i = 0\ , \quad h_i^B = - N_{\rm flux} \frac{S_i}{\S} \ , 
\end{align} 
admits an analytical solution for the real and imaginary parts of the axio-dilaton and all of the complex structure moduli, as long as the rest of the fluxes satisfy the constraints outlined above. Indeed, given such a choice of fluxes, one may compute the axionic  components using eqs.~\eqref{eq:bs}. On the other hand, we have seen that given the ansatz $t^i \equiv \hat{t} f_A^i$ for the complex structure saxions, one may use eq.~\eqref{eq:solthat} to compute $\hat{t}$ and, finally, use \eqref{eq:t0t} to determine the value of $t^0$. As a consequence, the search for flux vacua in the branch we have described here can be completely automatized.

Note that once $f_A^i$ and $h_i^B$ are fixed, one is free to choose $f_0^B$, $h_0^B$ and $f_i^B$ without changing the D3-tadpole. Thanks to the relation \eqref{eq:flux_const_away}, the definition of $\alpha$ \eqref{eq:def_alpha} and the definition of $Q'$ in \eqref{eq:rho}, these flux quanta may be easily tuned to generate vacua at the desired distance from the LCS point. This procedure has been explicitly carried out in the two-parameter example explored in section \ref{sec:numerics}.

In particular, this can also be useful to easily generate tuples of fluxes which yield vacua close to the LCS point, where exponentially suppressed 
corrections to the tree-level prepotential  may be neglected. From \eqref{eq:flux_const_away}, we find that vacua close to the LCS point where $|\xi| \ll 1$ satisfy
\begin{align}
	\label{eq:LCS_xi_flux}
	\xi \approx -  \frac{2^8}{3^5 (\Im \kappa_0)^2 \S} \, Q'^3 \ ,
\end{align}
where we recall that $\S\equiv \kappa_{ijk} f_A^i f_A^j f_A^k$ and $Q'$ has been defined in \eqref{eq:rho}. Thus, we need $Q'$ to be small and negative. An easy way to satisfy such a condition is by choosing $f_i^B = - f_A^j a_{ij}$, so that $L_i = 0$. In that case, $Q'$ is simplified to
\begin{align}
	f_i^B = - f_A^j a_{ij} \ \Longrightarrow \ Q' = f_0^B - f_A^i c_i + \frac{1}{2\S} \left( \frac{\S h_0^B}{N_{\rm flux}} \right)^2 .
\end{align}
Thus, having chosen $f_A^i$ and $N_{\rm flux}$, we can easily generate pairs of $f_0^B$ and $h_0^B$ which yield vacua with small $\xi$. 

\section{Supersymmetric vacua}
\label{sec:susy}

We now turn our attention to supersymmetric vacua which, as already mentioned, always contain a number of complex flat directions at the level of approximation to which we are working. One important feature of these vacua is that the flux quanta need to satisfy a series of constraints, in agreement with recent results in the literature. While obtaining the vevs for the stabilized fields is  straightforward, working out the mass spectra for these vacua turns out to be more involved than in the no scale aligned case.  

\subsection{Moduli stabilization and flat directions}

Here we describe the supersymmetric class of vacua defined in sect.~\ref{sec:Mnoninvertible}. As already said there and similarly to the case above, the requirement that $M\vec B=-\vec L$ generates $h^{2,1}+1-\rank (M)$ constraints that the fluxes must satisfy to fall into this case. The solutions for the moduli are expressed like $\vec Z=\vec B+\ker (M)$ such that there are $\rank (M)$ complex flat directions, and the additional requirement $W=0$ at vacua provides one more constraint on fluxes. If we put this back into the vacuum equations \eqref{eq:T_gen}, we obtain a simple linear system of equations where axions and saxions are decoupled:
\begin{align}
\begin{split}
 M\vec B&=-\vec L\ ,\\
M\vec T&=0\ .
\end{split}
\end{align}
The equation regarding the saxions can be further decomposed in the following relations
\begin{equation}
\label{eq:sijtj}
    h_i^B t^i=0\ ,\qquad S_{ij}t^j=h_i^Bt^0\ .
\end{equation}
Remembering now the decomposition discussed in \eqref{decomp}, we observe that supersymmetric vacua require $A=B=0$ and $C_i$, $C^i\neq0$, which contrasts with the set of non-supersymmetric solutions described by \eqref{eq:that}.

In order to make analytical progress, let us study again the subclass when the matrix $S$ possesses an inverse denoted $S^{ij}$ in components. When this is the case, then rank of $M$ is at least $h^{2,1}$ and for $M$ not to be invertible, it cannot be more than that. The non-invertibility of $M$ translates into the requirement
\begin{equation}
\label{eq:S0}
\cH=h_i^BS^{ij}h_j^B=0\ .    
\end{equation}
When solving $M\vec B=-\vec L$, as expected we derive one constraint and one axion is left unstabilized (this is the same situation as in sect.~\ref{sec:S0}):
\begin{align}
&h_i^BS^{ij}L_j=h_0^B\ ,\label{eq:cons_X}\\
&b^i=-S^{ij}L_j+b^0S^{ij}h_j^B\ .
\end{align}
Besides, the kernel of $M$ is one dimensional and given by
\begin{equation}
\ker (M)=\langle (1,S^{ij}h_j^B)\rangle\ .
\end{equation}
We thus have
\begin{equation}
\vec Z=\vec B+\ker (M)\quad\Longleftrightarrow\quad
\left\{
\begin{matrix*}[l]
\: \tau=b^0+\lambda\\[4pt]
z^i=b^i+\lambda S^{ij}h_j^B
\end{matrix*} 
\right. ,
\end{equation}
where $\lambda$ is some complex number that we can fix using the first equation of the system:
\begin{equation}
\Re (\lambda)=0\quad\text{ and }\quad \Im (\lambda)=t^0\ .    
\end{equation}
The second set of equations then gives expressions for $t^i$ with $t^0$ as a free parameter.\footnote{Note that here we applied naively the generic relation of sect.~\ref{sec:Bilinear} but we could have expressed $t^i$ easily from eq.~\eqref{eq:sijtj}.}  Summarizing, we have
\begin{align}
b^i&=-S^{ij}L_j+b^0S^{ij}h_j^B\ ,\label{eq:bi_susy}\\[5pt]
t^i&=S^{ij}h_j^Bt^0\ .
\end{align}
These relations define the two real flat directions that we expected from the general analysis.

One last constraint arising from the requirement of a vanishing superpotential is to be uncovered. Demanding $Q'=0$ from eq.~\eqref{eq:Q'} yields
\begin{equation}
\label{eq:cons_susy}
f_0-c_if_A^i-\half L_iS^{ij}L_j=0\ .
\end{equation}

Following similar arguments to the ones presented in section \ref{sec:generate_vacua}, a straightforward choice of fluxes which satisfy all the above conditions, eqs.~\eqref{eq:S0}, \eqref{eq:cons_X} and \eqref{eq:cons_susy}, is based on picking $f_A^i$ and $f^B_i$ such that
\begin{align}
	c_i f_A^i \in \mathbb{Z}\ , \quad a_{ij} f_A^j \in \mathbb{Z}\ , \quad f_i^B = - a_{ij} f_A^j \ .
\end{align}
This automatically implies
\begin{align}
	f_0^B = c_i f_A^i\ , \quad h_0^B = 0\ ,
\end{align}
so all that is left to do is to find $h_i^B$ such that
\begin{align}
	h_i^B  S^{ij} h_j^B = 0 \ .
	\label{eq:hhS}
\end{align}

Notice that the flux constraints \eqref{eq:S0} and \eqref{eq:cons_susy} agree with the tree-level conditions exposed in \cite{Demirtas:2019sip,Demirtas:2020ffz} where the authors further consider exponentially suppressed corrections in order to generate small flux superpotentials. The complex flat direction we found here when $S$ is invertible also seems to generalize the supersymmetric vacua uncovered in \cite{Cicoli:2022vny} to arbitrary Calabi--Yau geometries.

\subsection{Towards the mass spectrum}

In this section we push the computation of the mass spectrum for the supersymmetric vacua as far as we can. In the end, however, we will not be able to express it analytically in full generality like for the non-supersymmetric vacua with the simple saxionic ansatz. It is still interesting to understand what prevents us from doing so.

As we proved in the section above, the supersymmetric vacua satisfy
\begin{align}
	t^i = v^i t^0 \ , \quad v^i \equiv S^{ij} h_j^B\ ,\quad\text{ with } \quad h_i^B v^i = h_i^B S^{ij} h_j^B = 0\ .
\end{align}
We will follow the same logic as in the derivation of the mass spectrum for no-scale aligned vacua presented in appendix~\ref{sec:NSA}. This means we want to simplify the Kähler metric as best as we can, in order to obtain the simplest form possible for the matrix $Z_{AB}\equiv e^{K/2}D_AD_BW$ where the indices $A$, $B$ run into $\{\tau,z^i\}$. As reviewed in appendix \ref{sec:NSA} and shown in \cite{Sousa:2014qza}, the scalar masses $\mu_{\pm\lambda}$, $\lambda=0,\dots,h^{2,1}$ are simply given in the supersymmetric case by the fermion masses $m_{\lambda}$:
\begin{equation}
\mu_{\pm\lambda}=m_{\lambda}\ ,
\end{equation}
which correspond to the eigenvalues of the matrix $Z$.

To start orthonormalizing the Kähler metric \eqref{eq:Kij}, we can introduced two vielbeins inspired by the two preferred directions of the supersymmetric vacua: $t^i=t^0v^i$ and $f_A^i$. Notice, as we will explicitly see shortly, that in the non-supersymmetric branch studied earlier, these two vectors are aligned, which implies the alignment of $D_iD_\tau W$ with $K_i$ and hence the ``no-scale aligned'' property of the vacua, which enabled us to uncover the mass spectrum. We thus define the two vielbeins $e_1^i$ and $e_2^i$ like
\begin{equation}
    e_1^i\equiv \frac{t^i}{x}\quad\text{ and }\quad e_2^i\equiv \frac{f_A^i}{y}\ ,
\end{equation}
where $x$ and $y$ are normalization factors that can be straightforwardly expressed like
\begin{equation}
    x=\frac{\sqrt{3(2-\xi)}}{2(1+\xi)}\ ,\quad y=\sqrt{2t^0\Nf e^{K_{\rm cs}}}\ ,
\end{equation}
with $\Nf=-f_A^ih_i^B$. These two vielbeins are indeed orthogonal since we can show that
\begin{equation}
e_1^iK_{ij}e_2^j\propto h_i^BS^{ij}h_j^B=0\ .
\end{equation}
Plugging the vielbeins into the Kähler metric \eqref{eq:Kij}, we can obtain expressions for the rescaled Yukawa couplings $\mathring{\kappa}_{abc}$ involving the  direction $1$ similar to \eqref{eq:kappas_circle_app}:
\begin{align}
    \label{eq:kappas_circle}
    \mathring{\kappa}_{111} = \frac{2 (1+\xi)^2}{\sqrt{3(1-2\xi)^3}}\ ,\quad \mathring{\kappa}_{a' 11} = 0\ ,\quad \mathring{\kappa}_{a' b' 1} = \frac{-(1+\xi)}{\sqrt{3 (1 - 2 \xi)}} \delta_{a'b'}\ ,
\end{align}
where the prime indices run from $2$ onwards.

With this, we are now ready to see the special role played by these two directions: Direction $1$ is aligned with the no-scale direction while direction $2$ is aligned with $Z_{0a}$. Indeed, making use of \eqref{eq:Ki} and the symplectic decomposition of the flux vector \eqref{eq:N_structure} we find
\begin{equation}
    K_a=e_a^iK_i=2ix^2\mathring\kappa_{a11}\propto\delta_a^1\quad\text{ and }\quad Z_{0a}=ye^{K/2-K_{\rm cs}}\delta_a^2 \ .
\end{equation}
Finally, using eq.~\eqref{eq:Denef}, the expression for $Z_{ab}$ is
\begin{equation}
    Z_{ab}=-iye^{K/2-K_{\rm cs}}\mathring\kappa_{ab2}\ .
\end{equation}
Precisely because directions $1$ and $2$ are not aligned, we lack information to characterize the rescaled Yukawa couplings $\mathring\kappa_{ab2}$ and the only matrix elements we have control of are
\begin{equation}
\begin{aligned}
	Z_{0a} &= ye^{K/2-K_{\rm cs}}\delta_a^2\ ,\qquad\quad  &&Z_{11}=0\ , \\[7pt]
	Z_{1a} &= \frac{iy}{2x} e^{K/2-K_{\rm cs}} \delta_a^2\ , &&Z_{22} =- i e^{K/2} y^{-2} \S\ ,
\end{aligned}
\end{equation}
while the elements $Z_{2\tilde a}$ and $Z_{\tilde a\tilde b}$ are unknown for $\tilde a$, $\tilde b$ running from 3 onwards. The canonically normalized fermion mass matrix then reads
\begin{align}
	Z = \left(
	\begin{array}{ccc|c}
		0 & 0 & Z_{02} & 0  \\
		0 & 0 & Z_{12} & 0  \\
		Z_{02} & Z_{12} & Z_{22} & Z_{2\tilde a}  \\ \hline
		0 & 0 & Z_{2\tilde a} & Z_{\tilde a \tilde b} 
	\end{array}
	\right) .
	\label{eq:zab_w0}
\end{align}

Remember that the scalar masses correspond to the fermion ones, only doubled. The mass matrix \eqref{eq:zab_w0} cannot be diagonalized in full generality but it is easy to see that it features a massless mode, which thus translates into two massless directions in the scalar potential. This matches the expectations of the previous subsection.


\section{A numerical set of vacua in a two-parameter model}
\label{sec:numerics}

The goal of this section is to provide a numerical cross-check of the analytical results exposed in the previous section for the non-supersymmetric class of vacua following the \emph{no-scale aligned} branch with $t^i\propto f_A^i$. To this end, we generate an ensemble of IIB1 flux vacua in a two-parameter model by solving the vacuum equations numerically and then check various properties of these vacua. The model in question is the one arising from a symmetric point in the moduli space of the Calabi--Yau hypersurface $\mathbb{CP}^4_{[1,1,1,6,9]}$. We will first see how the analytical control of the IIB1 scenario enables us to generate a large number of vacua in the LCS regime very efficiently and we then show the perfect agreement between the features of these numerical vacua and the expectations from the analytics presented in sect.~\ref{sec:simpler}. 

\subsection{Generating flux tuples}

The first step to generate a numerical ensemble of vacua is to create a set of flux tuples meant to be run through in search for solutions of the vacuum equations. In order to reduce a bit the number of parameters, we consider the following restriction on the flux quanta $f_A^i$, $i=1,2$:
\begin{equation}
    f_A^1=f_A^2\equiv\hat f_A\ .
\end{equation}
If we trust our ansatz \eqref{eq:that}, this means that at the vacua we will have $t^1=t^2$.

We want flux configurations that do not overshoot the tadpole D3-charge bound $Q_{\text D3}$. With an O7-plane/D7-brane configuration identical to the one used in \cite{Demirtas:2019sip} and described in \cite{Louis:2012nb}, the induced D3-charge is restricted to satisfy $Q_{\text D3}\leq 138$. The flux contribution to the tadpole $\Nf$ depends only on $\hat f_A$ and $\hat h^B$ and thus we first generate a set of tuples for these flux quanta subject to the tadpole constraint. More precisely, we consider all flux entries in the range $[-6,6]$ and produce $14$ configurations satisfying the tadpole bound.

The fluxes remaining to be fixed at this point are $f_0^B$, $f_1^B$, $f_2^B$ and $h_0^B$. For the sake of efficiency, instead of generating a random set of tuples for them, we make use of our analytical expectations derived in sect.~\ref{sec:simpler}. This is done by expressing the flux-dependent quantity $\alpha$ defined in \eqref{eq:def_alpha} in terms of the unfixed flux quanta and by ensuring a choice of the latter such that $\alpha$ lies in the range $[-4,0]$. Since we want to cross-check our $\xi$-dependent analytics, we can do more than that and produce flux tuples that we expect to span the whole allowed range for $\xi$. To this end, we subdivide the $\alpha$ range $[-4,0]$ into $200$ pieces and try to find fluxes ${f_0^B,f_1^B,f_2^B,h_0^B}$ to fall into each piece, for each of the $14$ configurations ${\hat f_A,\hat h^B}$ previously generated. This results into a set of $2650$ full flux configurations that will use in the next subsection.\footnote{Note that all these steps are very easy and quick to implement so that a much bigger set of flux configurations could be generated effortlessly.}

\subsection{Vacua analysis}

\begin{figure}[!ht]
\centering
\includegraphics[scale=0.4]{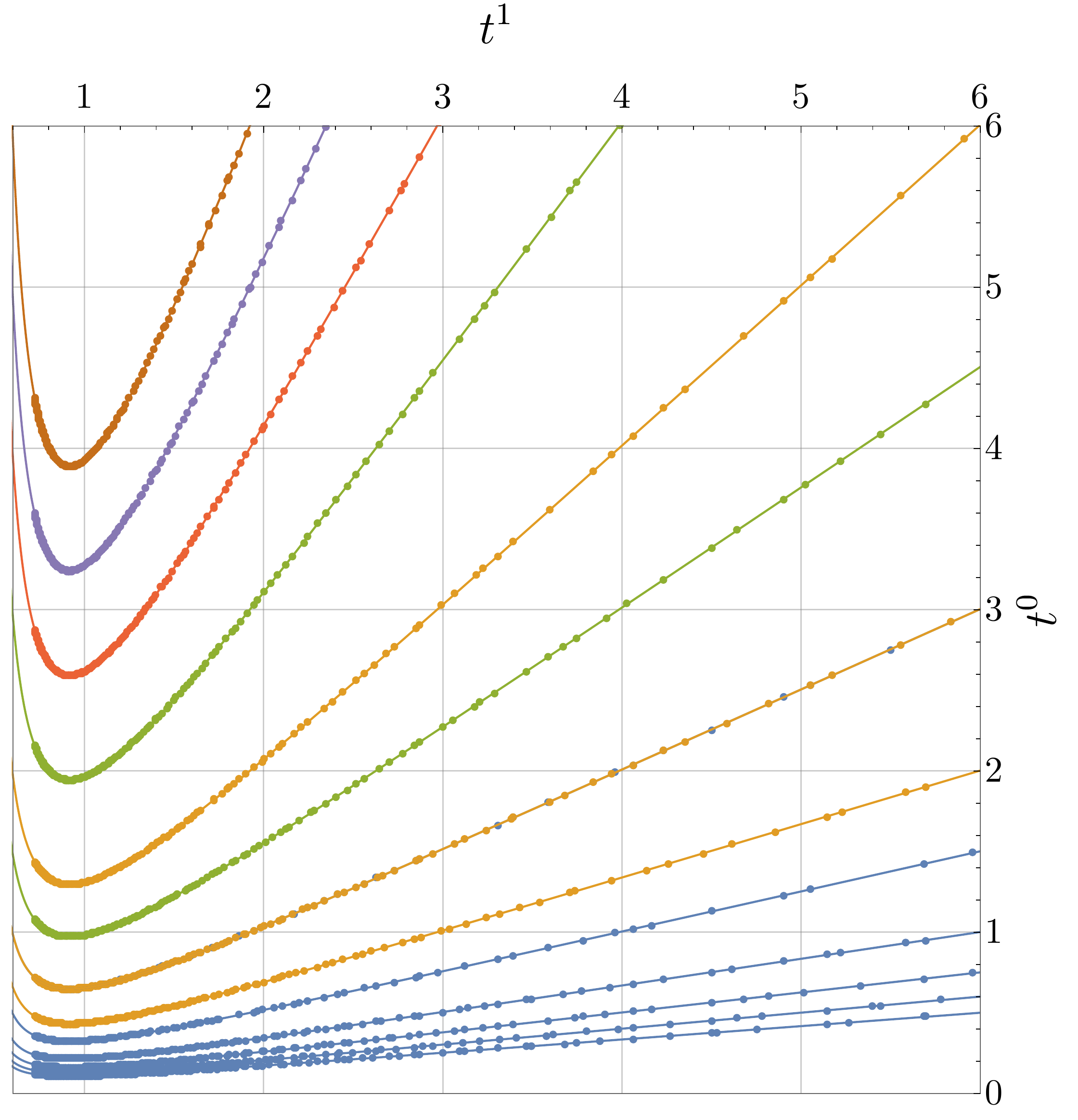}
\caption{This plot shows the locations of the numerically generated IIB1 vacua in the $(t^1,t^0)$-plane. The vacua are depicted with different colors corresponding to different values of $\hat f_A$, with different branches corresponding to different values for $\hat h^B$, present in the ensemble. For a given color, the expression \eqref{eq:t0t_num} is displayed on top of the numerical data. We observe a perfect agreement. More precisely, the colors correspond to the following fluxes: Blue: $\hat f_A=1,\,\hat h^B=1,\dots,6$; Orange: $\hat f_A=2,\,\hat h^B=1,\dots,3$; Green: $\hat f_A=3,\,\hat h^B=1,2$; Red: $\hat f_A=4,\,\hat h^B=1$; Purple: $\hat f_A=5,\,\hat h^B=1$; and Brown: $\hat f_A=6,\,\hat h^B=1$. As explained in sect.~\ref{sec:instantons}, vacua with $\xi<0.17$ \ie with $t^1,\ t^2\gtrsim 1$ are expected to be safe under instanton corrections as the relative changes induced by the corrections on the moduli space and other quantities are small.}
\label{t1t0_fit}
\end{figure}

We numerically implemented the vacuum equations and searched for solutions for each flux configuration of our ensemble. The two-parameter model is characterized by the following topological quantities that fully define the prepotential \eqref{eq:full_prepotential} (neglecting exponentially suppressed corrections):
\begin{equation}
\begin{aligned}
&\kappa_{111}=9\ ,\quad&&\kappa_{112}=3\ ,\quad &&\kappa_{122}=1\ ,\quad\kappa_{222}=0\ ,\\
&\kappa_{11}=-\frac{9}{2}\ ,\quad&&\kappa_{22}=0\ ,\quad&&\kappa_{12}=-\frac{3}{2}\ ,\\
&\kappa_1=\frac{17}{4}\ ,\quad&&\kappa_2=\frac{3}{2}\ ,\quad&&\kappa_0=-540\frac{\zeta(3)}{(2i\pi)^3}\ .
\end{aligned}
\end{equation}
As expected from our careful choice of fluxes guided by the analytics, each flux tuple yields a consistent vacuum inside the Kähler cone. The vacua are displayed in the $(t^1,t^0)$-plane in fig.~\ref{t1t0_fit}.

A first analytical relation that we can check is eq.~\eqref{eq:t0t}. In the case at hand with $f_A^1=f_A^2=\hat f_A$, we have $q=(\hat f_A)^2$ and $\S=21(\hat f_A)^3$. The relation then becomes 
\begin{equation}
\label{eq:t0t_num}
t^0 =-\frac{\hat f_A}{\hat h^B}\frac{14 (t^1)^3-\Im\kappa_0}{28(t^1)^3+\Im\kappa_0} \ t^1\ .
\end{equation}
The comparison between this analytical formula and the data of our ensemble of vacua is displayed in fig.~\ref{t1t0_fit}. We observe a perfect match between the two.

\begin{figure}[!ht]
\centering
\includegraphics[scale=0.48]{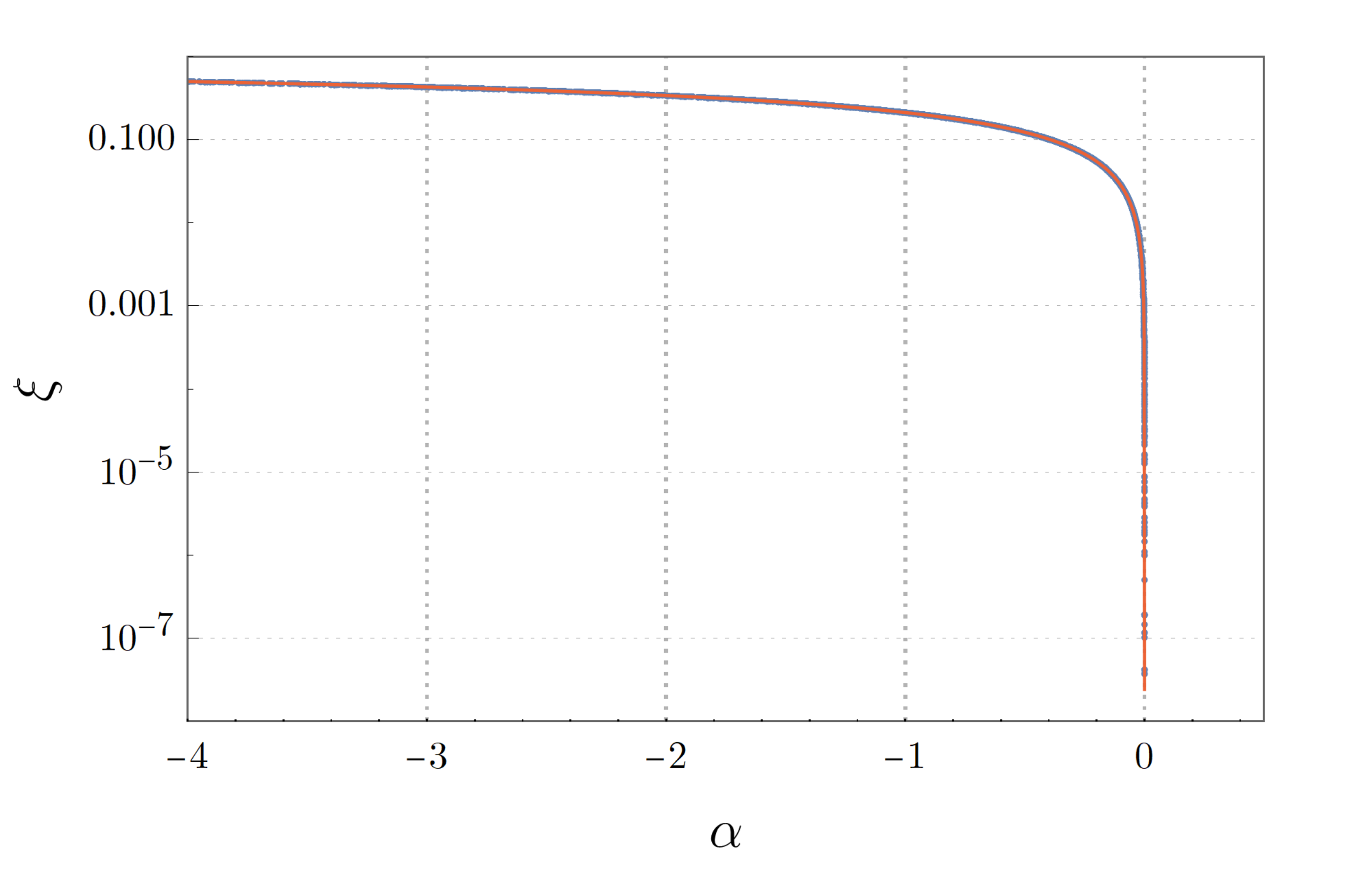}
\caption{This plots shows the values of $\xi$ against $\alpha$ for the numerical vacua of our ensemble. The relation \eqref{eq:flux_const_away} is plotted in red and fits perfectly the data points.}
\label{xi_alpha}
\end{figure}

Another non-trivial result we can check is the relation between $\xi$ and the quantity $\alpha$ (see eqs.~\eqref{eq:def_alpha} and \eqref{eq:flux_const_away}). Figure \ref{xi_alpha} shows a nice fit of the data by the analytical expression.

\begin{figure}[!ht]
\centering
\includegraphics[scale=0.5]{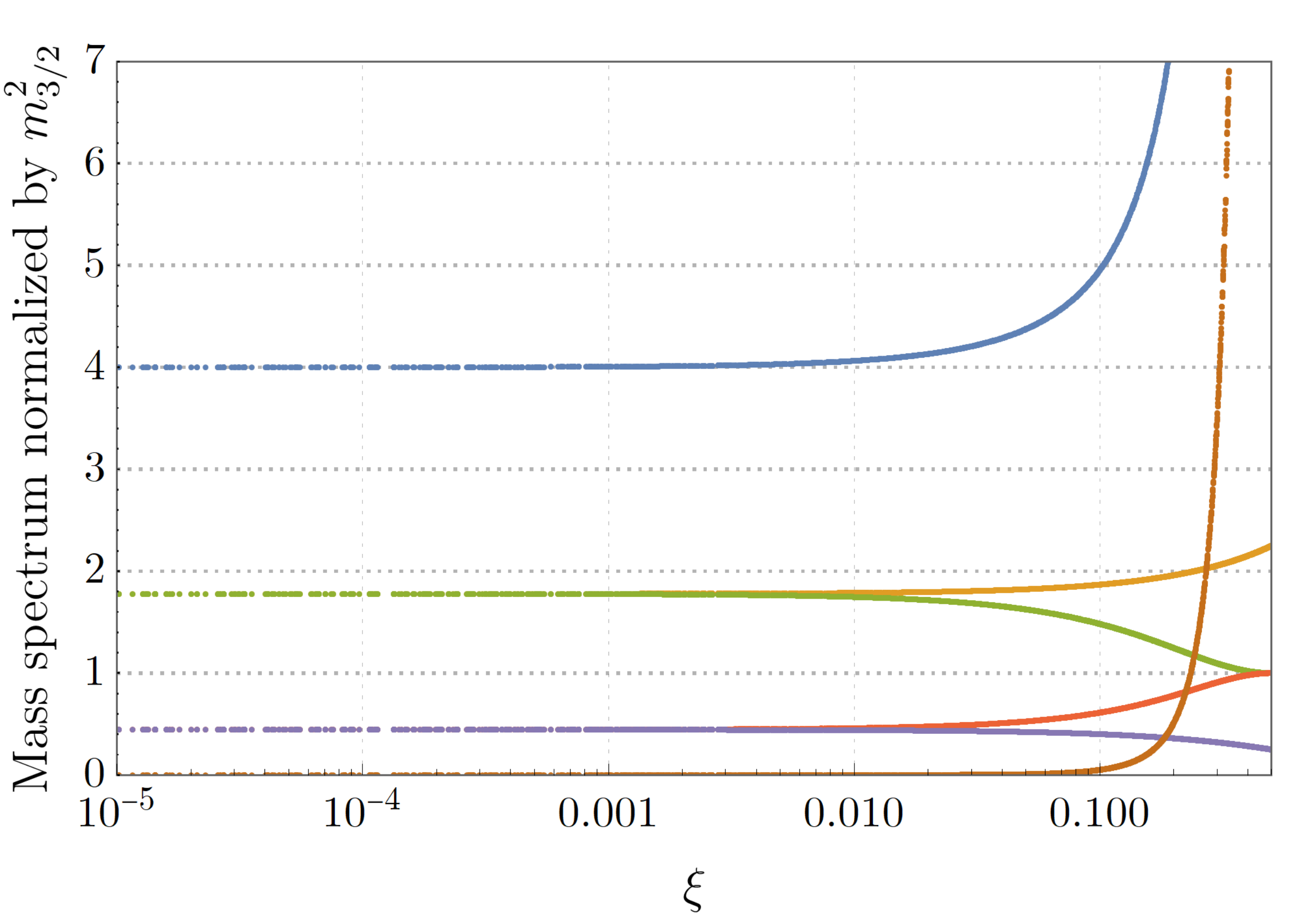}
\caption{This plot shows the squared masses, normalized by the gravitino mass squared, numerically obtained in the set of vacua. They precisely reproduce the analytical behaviour \eqref{eq:nsa_mass_spectrum} displayed in fig.~\ref{analytical_mass_spectrum}.}
\label{spectrum}
\end{figure}

One last important result to be checked is the mass spectrum in the vacua. We have shown in sect.~\ref{sec:simpler} that the vacua under consideration fall into the definition of the \emph{no-scale aligned} setup whose mass spectrum normalized by the gravitino mass $m_{3/2}$ is given as a function of $\xi$ by eq.~\eqref{eq:nsa_mass_spectrum}. The canonically normalized masses, numerically computed for each vacuum, are displayed in fig.~\ref{spectrum}. We again observe that the numerical results perfectly match the analytical expectations displayed in fig.~\ref{analytical_mass_spectrum} in sect.~\ref{sec:masses}.

\subsection{Exponential corrections}
\label{sec:instantons}

Of course we expect exponential corrections in the prepotential \eqref{eq:full_prepotential} to become more and more relevant as the LCS parameter $\xi$ goes away from the LCS point and gets closer to the boundary at $\xi=1/2$. In specific examples and following \cite{Blanco-Pillado:2020wjn,Blanco-Pillado:2020hbw}, we can evaluate the effect of the exponentially suppressed corrections by computing their relative effects on the geometry of the moduli space and other physical quantities.

For the $\mathbb{CP}^4_{[1,1,1,6,9]}$ hypersurface, the dominant exponential corrections are expressed like \cite{Blanco-Pillado:2020hbw,Candelas:1994hw}
\begin{equation}
\F_{\rm inst}=-\frac{135}{2\pi^3}ie^{2i\pi z^1}-\frac{3}{8\pi^3}ie^{2i\pi z^2}\ ,
\end{equation}
and we can use them to numerically compute the relative errors induced on the Kähler metric, the gravitino mass $m_{3/2}$ and $e^{K_{\rm cs}}$. Note that this definition for the validity of the perturbative result is rather conservative and much more stringent than only requiring the non-perturbative part of the prepotential to be dominated by the perturbative one. We find that vacuum expectation values for $t^1=t^2$ slightly above $1$ are enough to guarantee the stability of the perturbative vacua since all the relative corrections are smaller than a small threshold of $5\%$. In terms of the LCS parameter, $\xi<0.17$ ensures robustness of the perturbative results.


\section{Conclusions and outlook}
\label{sec:Conclusion}

In this paper, we investigated a specific type IIB family of flux vacua at large complex structure introduced in \cite{Marchesano:2021gyv} and called \emph{IIB1 scenario}. Arising as a type IIB limit from an F-theory construction, the vacuum equations were studied there at first order in the LCS parameter $\xi$ defined in \eqref{eq:def_xi}, i.e., not too far from the LCS point. Our analysis extends these results by exploring in more detail different classes of vacua allowed by the IIB1 setup, and by pushing their analytical resolutions (computation of the complex structure and axio-dilaton vevs as well as mass spectra) as far as possible.

The IIB1 choice of fluxes ensures that all cubic terms disappear from the flux-induced superpotential such that it is simply quadratic in the axio-dilaton and complex structure fields. A very generic and coarse-grained classification of vacua arising from such a quadratic structure reveals the existence of one supersymmetric family and two non-supersymmetric ones, depending on the definiteness or not of the bilinear form involved in the superpotential. More precisely, a regular bilinear structure forbids supersymmetric vacua while a singular one allows vacua that are either supersymmetric or not. In any of these cases, the vacuum equations nicely split into two separate systems: A very simple one involving only the axions (thanks to the independence on the axions of the superpotential at vacua), and a more involved one relating the saxions. Moduli stabilization can then be studied separately for these two sets of fields.

We then explored the three classes mentioned above further in detail. The supersymmetric vacua are described by very simple vacuum equations thanks to the vanishing of the superpotential on-shell. Restricting to fluxes such that the matrix $S$, with $S_{ij}\equiv \kappa_{ijk}f_A^k$ involving the triple intersection numbers of the mirror manifold, is invertible (a recurring assumption in this paper), we saw that the supersymmetric vacua feature one complex flat direction and are similar to those used in \cite{Demirtas:2019sip,Demirtas:2020ffz} to achieve small superpotentials. They also generalize the supersymmetric models studied in \cite{Cicoli:2022vny} to arbitrary Calabi--Yau compactifications.  For these supersymmetric vacua, we addressed the computation of the  scalar masses, and it seems that further analytical progress in obtaining the mass spectrum for models with $h^{2,1}>2$ requires more definite knowledge of the model under study.

The two non-supersymmetric classes highlighted above differ if the bilinear form involved in the superpotential is degenerate or not. The effect of a non-trivial kernel is to generate one flux constraint and one flat direction for each dimension of the kernel of the bilinear form. As a particular case, when the matrix $M$ representing the form is invertible, all axions are stabilized. Whether $M$ is regular or not, the saxionic system of equations is highly non-linear and generically stabilizes all fields. As a counterpart, it is trickier to handle. To make analytical progress, we proposed an ansatz \eqref{eq:ansatz} for the saxions and studied the subsequent vacuum equations. This led us to consider two further refined branches where we could provide analytic expressions for all the vevs of the axio-dilaton and complex structure fields, and even express analytically the scalar mass spectrum for one of these branches.

The first branch is a subcase where the matrix $M$ is singular with a specific uni-dimensional kernel. One axionic direction is thus left as a flat direction. The saxionic vacuum equations produce a sixth order polynomial relation, from which we can express the saxion vevs. The polynomial can be analytically solved using a perturbative expansion in the LCS parameter $\xi$. The second branch is uncovered when assuming a simpler sub-ansatz \eqref{eq:that} for the saxions. It is shown to be allowed only when $M$ is regular, so that all axions are fixed. The saxionic system yields a manageable cubic polynomial such that the vevs can be fully expressed within the LCS region. Moreover, we showed that this branch falls into the \emph{no-scale aligned} family studied in \cite{Blanco-Pillado:2020wjn,Blanco-Pillado:2020hbw}, for which the scalar mass spectrum can be fully expressed analytically in terms of the LCS parameter. As already observed in \cite{Marchesano:2021gyv} and expected from the necessity of incorporating polynomial corrections to stabilize all moduli in this context, these kind of mass spectra feature a mode becoming lighter  as one gets closer to the LCS point.

We checked numerically the validity  of our  approximations in the non-supersymmetric \emph{no-scale aligned} branch, and in particular the accuracy of the mass spectrum. We did this by investigating a small ensemble of IIB1 vacua in this branch, generated numerically. We worked with the two-parameter model coming from a symmetric point in the moduli space of the Calabi--Yau hypersurface $\mathbb{CP}^4_{[1,1,1,6,9]}$. In addition to providing a solid cross-check of the analytics derived in the paper, the numerical analysis shows that the IIB1 scenario provides a setup where we can very efficiently generate vacua numerically at (almost) arbitrary distance of the LCS point desired. We can compactly summarize the analytical results for this highly controllable branch of vacua as follows:

\subsection{Summary of analytic type IIB mass spectra}

If one considers a Calabi--Yau orientifold of IIB string theory described by the tree-level prepotential
\begin{equation}
		\mathcal{F}=-\frac{1}{6}\kappa_{ijk}z^iz^jz^k-\frac{1}{2}a_{ij}z^iz^j+c_iz^i+\frac{1}{2} \kappa_0\ ,
\end{equation}
and with fluxes subject to
\begin{align}	
	f_A^0=0\ , \quad h_A^0=0\ , \quad h_A^i = 0\ , \quad h_i^B = - q^{-1} \hat{h}^B \kappa_{ijk} f_A^j f_A^k \ ,
\end{align}
where $\hat{h}^B \in \mathbb{Z}$ and $q\equiv\gcd (\kappa_{ijk} f_A^j f_A^k)$. Then, there exist non-supersymmetric no-scale vacua, i.e. configurations of the moduli that satisfy $D_{\tau} W = D_i W = 0$ for the axio-dilaton $\tau \equiv b^0 + i t^0$ and the complex structure moduli $z^i \equiv b^i + i t^i$, such that all moduli are stabilized at
\begin{equation}
\begin{aligned}
		&b^0 = - \dfrac{q}{\hat{h}^B \S} \left[ f_A^i L_i+ \dfrac{q h_0^B}{\hat{h}^B} \right]\ ,\qquad	&&t^0 = 	\left( \dfrac{3 | \Im \kappa_0 |}{2\S} \right)^{1/3} \dfrac{1+\xi}{q^{-1} \hat{h}^B (2-\xi) \xi^{1/3}}\ , \\[25pt]
		&b^i = \dfrac{1}{\S} \left[ f_A^j L_j + \dfrac{q h_0^B}{\hat{h}^B} \right] f_A^i - S^{il} L_l\ ,\qquad &&t^i = \left( \dfrac{3 | \Im \kappa_0 |}{2\S\xi} \right)^{1/3} f_A^i\ ,
\end{aligned}
\end{equation}
where we have defined $L_i \equiv f_i^B + a_{ij}f_A^j$, $S_{ij} \equiv \kappa_{ijk} f_A^k$ and its inverse $S^{ij}$, and $\S \equiv \kappa_{ijk} f_A^i f_A^j f_A^k$. In the expressions above, $\xi$ parameterizes the distance of the vacuum to the LCS point located at $\xi=0$. It can also be written entirely in terms of fluxes as
\begin{align}
	\xi=2+\frac{9+\alpha\delta^2}{\alpha\delta}
\end{align}
where 
\begin{align}
\begin{split}
		&\alpha\equiv \frac{32 }{9|\Im\kappa_0|^2 \S} \left( f_0^B - c_i f_A^i + \frac{1}{2\S} \left[ f_A^i L_i+ \frac{q h_0^B}{\hat{h}^B} \right] - \frac{1}{2} L_i S^{ij} L_j \right)^3 \ , \\[5pt]
		&\delta\equiv3\left[\alpha^2+\sqrt{\alpha^3(\alpha-1)}\right]^{-1/3}\ .
		\label{eq:def_alpha_beta_con}
\end{split}		
\end{align}
Furthermore, these vacua correspond to the \emph{no-scale aligned} class, which implies their scalar mass spectrum $\mu^2_{\pm \lambda}$, normalized by the gravitino mass squared $m_{3/2}^2\equiv e^{K} |W|^2 $, can be written analytically as
\begin{equation}
	\frac{\mu^2_{\pm \lambda}}{m_{3/2}^2} = 
	\left\lbrace
	\begin{array}{ll}
		\left( 1 \pm \sqrt{\frac{1 -2 \xi}{3}} \hat{m} (\xi) \right)^2  & \lambda = 0  \\
		\left( 1 \pm \sqrt{\frac{1 -2 \xi}{3}} (\hat{m} (\xi))^{-1} \right)^2  & \lambda = 1 \\
		\left( 1 \pm \frac{1+\xi}{3} \right)^2 & \lambda = 2, \ldots, h^{2,1}
	\end{array}
	\right.
	\label{eq:nsa_mass_spectrum_con}
\end{equation}
where we have defined
\begin{align}
\begin{split}
&\hat{m} (\xi) \equiv \frac{1}{\sqrt{2}} \left( 2 + \kappa (\xi)^2 - \kappa (\xi) \sqrt{4 + \kappa (\xi)^2} \right)^{1/2}\ ,\\
&\kappa (\xi)\equiv 2 (1+ \xi)^2 / \sqrt{3(1-2\xi)^3}\ .
\end{split}    
\end{align}

\subsection{Outlook}

Our simple ansatz presented in sect.~\ref{sec:simpler} allows for complete analytical control over both the distance to the LCS point and the vevs of all complex structure moduli and the axio-dilaton. As such, this setup can be extremely useful to consider further corrections to the tree-level solutions, either by the inclusion of stringy corrections which would render more accurate solutions, or by the inclusion of exponentially suppressed corrections to the prepotential. An interesting line of work in this sense can be the stabilization of the Kähler sector through different means, either through racetrack potentials \cite{Kachru:2003aw,Balasubramanian:2005zx} or by more generic mechanisms 
\cite{AbdusSalam:2020ywo}.

The analytics derived in this paper hold for models with an arbitrary number of complex structure moduli at large complex structure. However, one should keep in mind that when the number of moduli is large, the flux-induced contribution to the D3-brane tadpole may go out of control as proposed by the Tadpole Conjecture \cite{Bena:2020xrh,Bena:2021wyr}. In the setup of our simple ansatz, it is worth noticing that our estimates for the flux-induced tadpole $\Nf$ are in the same footing as the solutions discussed in \cite{Plauschinn:2021hkp,Lust:2021xds,Grimm:2021ckh,Grana:2022dfw}. This is because, on the one hand, the ansatz forces the flux quanta $f_A^i$ to be non-zero and to have a same common sign for the saxionic vevs to be well-defined. On the other hand, the constraint \eqref{hBicond} on the fluxes $h_i^B$ also imposes these quanta to be non-zero and have the same sign, such that $\Nf$ is a generically a sum of $h^{2,1}$ positive terms \cite{Marchesano:2021gyv}. As a consequence, the tadpole contribution indeed grows with the number of moduli in this context. However, we cannot say much more in this sense for the more involved ansatz \eqref{eq:ansatz} where flux quanta are less restricted or even for solutions outside this generic ansatz.

We should also point out that in our numerical analysis we are using a model where effectively only two moduli play the game thanks to a consistent truncation, and thus, small tadpoles can be achieved there without too much tinkering. This is also in line with \cite{Lust:2022mhk}, where a similar reasoning is applied to F-theory compactifications built at loci of discrete symmetry groups of the moduli space. Even though the tadpole conjecture is generically very sound, it is also true that such symmetric models may allow for non-generic solutions where the tadpole is small. We expect to answer such claims in the large complex structure regime of type IIB string theory in a future work.


\section*{Acknowledgments}

We thank Jose Juan Blanco-Pillado for discussions. This work is supported through the grants EUREXCEL$\_$03 funded by CSIC, CEX2020-001007-S and PID2021-123017NB-I00, funded by MCIN/AEI/10.13039/501100011033 and by ERDF A way of making Europe, and by the Spanish Ministry MCIU/AEI/FEDER 
grant (PID2021-123703NB-C21). D. P. is supported through the grant FPU19/04298 funded by MCIN/AEI/10.1\-3039/501100011033 and by ESF Investing in your future. 


\appendix
\makeatletter
\DeclareRobustCommand{\@seccntformat}[1]{%
  \def\temp@@a{#1}%
  \def\temp@@b{section}%
  \ifx\temp@@a\temp@@b
  \appendixname\ \thesection:\quad%
  \else
  \csname the#1\endcsname\quad%
  \fi
} 
\makeatother
\renewcommand{\theequation}{A.\arabic{equation}}

\section{Mass spectrum of no-scale aligned vacua}
\label{sec:NSA}

The no-scale aligned vacua described in \cite{Blanco-Pillado:2020wjn} are defined by the following relation between two-derivatives of the superpotential and one-derivative of the Kähler potential as well as two constraints on flux quanta:
\begin{align}
	D_{\tau} D_i W \propto K_i\quad\text{ and }\quad f_A^0=h_A^0=0\ .
	\label{eq:nsa_def}
\end{align}
These vacua feature an analytical mass spectrum expressed solely in terms of the LCS parameter. In this section, we present the key steps of its derivation.

One of the main difficulties to obtain the mass spectrum in generic points of field-space is the fact that one has to compute eigenvalues with respect to the field space metric $K_{i\bar{j}}$. In order to overcome this difficulty, it is customary to introduce real vielbein $e_i^a$ which render the metric to a canonical form \cite{Blanco-Pillado:2020wjn,Marsh:2015zoa}, such that
\begin{align}
	K_{i\bar{j}} = e_i^a \delta_{ab} e_{\bar{j}}^b \ , \quad \delta_{ab} = e_a^i K_{i\bar{j}} e_b^{\bar{j}} \ .
\end{align} 
In what follows, we will reserve letters $i,j,\ldots$ to refer to curved indices in field space, while $a,b,\ldots$ will label flat indices. 

From the block-diagonal form of the metric in the axio-dilaton and complex structure sectors, we can easily see that $e_0^{\tau} = -2 t^0$. On the other hand, since we are free to choose the first vielbein to diagonalize the metric, we will pick 
\begin{align}
	e_1^i \equiv \frac{t^i}{x}\ ,
\end{align}
where $x$ is a normalization factor. Plugging this into the field-space metric \eqref{eq:Kij}, we get
\begin{align}
  \delta_{ab} = e_a^i K_{i\bar{j}} e^j_b = - 2 x \mathring{\kappa}_{ab1} + 4 x^4 \mathring{\kappa}_{a11} \mathring{\kappa}_{b11}\ .
\end{align}
Using the definition of the LCS parameter \eqref{eq:def_xi} and the previous equation, we can obtain several identities:
\begin{align}
    \label{eq:kappas_circle_app}
	x = \frac{\sqrt{3 (1 -2 \xi)}}{2(1 + \xi)}\ , \ \mathring{\kappa}_{111} = \frac{2 (1+\xi)^2}{\sqrt{3(1-2\xi)^3}}\ , \ \mathring{\kappa}_{a' 11} = 0\ , \ \mathring{\kappa}_{a' b' 1} = \frac{-(1+\xi)}{\sqrt{3 (1 - 2 \xi)}} \delta_{a'b'}\ ,
\end{align}
where the prime indices $a',b'$ run from 2 onwards. 

The ``1'' direction in the vielbein turns out to have special significance. Contracting its corresponding vielbein with $K_i$ given in eq.~\eqref{eq:Ki}, we find
\begin{align}
	Z_{0a}\propto K_a = e_a^i K_i = 2 i x^2 \mathring{\kappa}_{a11} \propto \delta_a^1\ ,
\end{align}
where the matrix $Z$ is defined below. Thus, the introduction of the vielbein into our problem not only simplifies expressions involving the field-space metric or its inverse, but it also aligns the so-called no-scale direction $K_i$ with the 1-direction. 

We are now prepared to tackle the computation of the mass spectrum. In order to do this, we will proceed in the lines of \cite{Blanco-Pillado:2020wjn}. As explicitly proven in that work, the mass spectrum can be neatly written as \cite{Sousa:2014qza}
\begin{equation}
	\mu^2_{\pm \lambda} = (m_{3/2} \pm m_{\lambda})^2\ ,
	\label{eq:mass_rel}
\end{equation}
where $\lambda = 0, \ldots, h^{2,1}$ and we have defined the gravitino mass $m_{3/2} \equiv e^{K/2} |W|$ as well as the fermion masses $m_{\lambda}$. The easiest way to obtain the latter ones is through the diagonalization of the following matrix\footnote{The first metric factor must be introduced due to the kinetic term of the scalar fields being potentially non-canonical.}
\begin{align}
	(Z^{\dag} Z)^A_B \equiv K^{A\bar{C}} \bar{Z}_{\bar{C}\bar{D}} K^{\bar{D}E} Z_{EB}\ , 
\end{align}
where $Z_{AB} \equiv e^{K/2} D_A D_B W$ and the indices $A,B,\dots$ run into $\{\tau,z^i\}$. Thus, eigenvalues of $Z^{\dag}Z$ will yield the masses  $m^2_{\lambda}$. 

In order to compute these values we will employ several simplifying schemes. First of all, it is easy to check that $Z_{\tau \tau} = 0$ at supersymmetric vacua described by the tree-level LCS prepotential. Another useful identity is \cite{Denef:2004ze,Candelas:1990pi}
\begin{equation}
\label{eq:Denef}
	Z_{ij} = - (\tau - \bar{\tau}) e^{K_{\rm cs}} \kappa_{ijk} K^{k\bar{l}} \bar{Z}_{\bar{\tau} \bar{l}}\ .
\end{equation}
This identity can be easily rewritten in terms of the vielbein introduced above:
\begin{align}
	Z_{ab} = i \mathring{\kappa}_{abc} \delta^{cd} \bar{Z}_{0d}\ .
	\label{eq:zab_id}
\end{align}
On the other hand, since the vacua we are studying have the no-scale-aligned property \eqref{eq:nsa_def}, we have that $Z_{0a} \propto \delta_a^1$ and therefore,
\begin{align}
	Z_{ab} = i \mathring{\kappa}_{ab1} \bar{Z}_{01} \ . 
\end{align}
Note that we have a closed expression in terms of $\xi$ for all the required $\mathring{\kappa}_{ab1}$ that will appear when constructing $Z$. Using eq.~\eqref{eq:kappas_circle_app}, the matrix reads
\begin{equation}
Z_{AB}\!=\!\begin{pmatrix}
0 & Z_{01} & 0\\
Z_{01} & i\mathring{\kappa}_{111}\bar Z_{01} & 0\\
0 & 0 & i\mathring{\kappa}_{a'b'1}\bar Z_{01}
\end{pmatrix}
\!=\!\begin{pmatrix}
0 & Z_{01} & 0\\
Z_{01} & \frac{2i (1+\xi)^2}{\sqrt{3(1-2\xi)^3}}\bar Z_{01} & 0\\
0 & 0 & \frac{-i(1+\xi)}{\sqrt{3(1-2\xi)}}\delta_{a'b'}\bar Z_{01}
\end{pmatrix}.
\end{equation}

The diagonalization of $Z^\dagger Z$ gives the following eigenvalues:
\begin{equation}
	m_\lambda^2 = 
	\left\lbrace
	\begin{array}{ll}
		 \hat{m} (\xi)^2|Z_{01}|^2  & \lambda = 0  \\[5pt]
		(\hat{m} (\xi))^{-2}|Z_{01}|^2  & \lambda = 1 \\[5pt]
		\frac{(1+\xi)^2}{3(1-2\xi)}|Z_{01}|^2 & \lambda = 2, \ldots, h^{2,1}
	\end{array}
	\right.
	\label{eq:fermion_masses}
\end{equation}
where we have defined the quantities
\begin{align}
\begin{split}
&\hat{m} (\xi) \equiv \frac{1}{\sqrt{2}} \left( 2 + \kappa (\xi)^2 - \kappa (\xi) \sqrt{4 + \kappa (\xi)^2} \right)^{1/2}\ ,\\
&\kappa (\xi)\equiv\mathring{\kappa}_{111}= 2 (1+ \xi)^2 / \sqrt{3(1-2\xi)^3}\ .
\end{split}    
\end{align}
In order to deal with the dependency on $|Z_{01}|$, we use the other defining feature of no-scale aligned vacua, namely $f_A^0 = h_A^0 = 0$. According to the decomposition of the flux vector given in eq.~\eqref{eq:N_structure} together with the form of the period vector \eqref{eq:period}, this choice of fluxes leads to
\begin{align}
	W = -2i t^0 D_{\bar{\tau} \bar{j}} \bar{W} K^{\bar{j}i} K_i = i D_{0a} W \delta^{ab} K_b  \ \Rightarrow \ &
	& m_{3/2} = \sqrt{\frac{3}{1 - 2 \xi}} |Z_{01}| .
	\label{eq:m32_id}
\end{align}
Therefore, when plugging the eigenvalues $m_{\lambda}^2$ into eq.~\eqref{eq:mass_rel}, we can factorize an $m_{3/2}^2$ factor and obtain the scalar masss spectrum at no-scale-aligned vacua:
\begin{equation}
    \text{NSA mass spectrum:}\quad
	\frac{\mu^2_{\pm \lambda}}{m_{3/2}^2} = 
	\left\lbrace
	\begin{array}{ll}
		\left( 1 \pm \sqrt{\frac{1 -2 \xi}{3}} \hat{m} (\xi) \right)^2  & \lambda = 0  \\
		\left( 1 \pm \sqrt{\frac{1 -2 \xi}{3}} (\hat{m} (\xi))^{-1} \right)^2  & \lambda = 1 \\
		\left( 1 \pm \frac{1+\xi}{3} \right)^2 & \lambda = 2, \ldots, h^{2,1}
	\end{array}
	\right.
	\label{eq:nsa_mass_spectrum_app}
\end{equation}

\renewcommand{\theequation}{B.\arabic{equation}}

\section{Scalar potential and mass matrix}
\label{ap:spot}

In this section we present a detailed derivation of the scalar potential that describes the  IIB1 scenario and use it to directly compute the Hessian of the axionic sector, hence providing an alternative way to obtain the associated mass spectrum. 

\subsection{Metric tensor}

In the main text we found the vacuum equations using the no-scale structure of type IIB and working only with the superpotential. This procedure proved to be a powerful simplifying tool. However we now wish to go back to the results of \cite{Marchesano:2021gyv} and write the scalar potential for the  IIB1 scenario with corrections to all orders. The first step in this process is to revisit the Kähler potential and analyse the moduli space metric in more detail. From \eqref{eq: complex Kahler potential} we have
\begin{equation}
     K_{\text{cs}} = - \log \left( \frac{4}{3} \kappa_{ijk} t^i t^j t^k (1+\xi) \right)\ .
\end{equation}
Taking partial derivatives with respect to the dilaton and complex structure moduli we find
 \begin{align}
    K_{\tau \tau}=&\ \frac{1}{4(t^0)^2}\ ,\\
    K_{\tau i}=&\ 0\ ,\\
    K_{ij}=&\ \frac{9}{4}\frac{\kappa_i\kappa_j}{\kappa^2(1+\xi)^2}-\frac{3}{2}\frac{\kappa_{ij}}{\kappa(1+\xi)}=K^{\rm o}_{ij}\frac{\kappa}{\kappa(1+\xi)}-\frac{9}{4}\frac{\kappa_i\kappa_j}{\kappa^2(1+\xi)^2}\xi\ ,
\end{align}  
with $\kappa_{ij}\equiv\kappa_{ijk} t^k$, $\kappa_i\equiv\kappa_{ij}t^j$ and $\kappa\equiv\kappa_i t^i$. Finally we denote by $K^{\rm o}_{ij}$ the leading order metric, that is, the metric in the limit $\xi\rightarrow 0$.

Using the last expression for $K_{ij}$, it is straightforward to obtain its inverse in terms of the inverse of the leading order metric:
\begin{equation}
    K^{ij}=K_{\rm o}^{ij}(1+\xi)+\frac{4\xi(1+\xi)}{1-2\xi}t^it^j\ .
\end{equation}
Following the same line of reasoning as in \cite[Appendix B.3]{Marchesano:2021gyv} we also compute
\begin{equation}
    K^{ij}K_{j}=2it^i\frac{1+\xi}{1-2\xi}\ .
\end{equation}
Finally, note that the metric leading order metric splits in its primitive and non primitive components as
\begin{align}
\begin{split}
    K_{ij}^{\rm oNP}&=\frac{3}{4}\frac{\kappa_i\kappa_j}{\kappa^2}\ ,\\
    K_{ij}^{\rm oP}&=\frac{3}{2}\frac{\kappa_i\kappa_j}{\kappa^2}-\frac{3}{2}\frac{\kappa_{ij}}{\kappa}\ .
\end{split}
\end{align}
In particular they satisfy $K_{ij}^{\rm oP}t^j=0$ and $K^{ij}_{\rm oP}\kappa_j=0$. We can replicate this split for the full metric to find
\begin{align}
\begin{split}
    K_{ij}^{\rm NP}&=K_{ij}^{\rm oNP}\frac{1-2\xi}{(1+\xi)^2}\ ,\\
    K_{ij}^{\rm P}&=K_{ij}^{\rm oP}\frac{1}{1+\xi}\ .
    \label{eq: primitive and no primitive corrected}
\end{split}
\end{align} 
\subsection{Scalar potential}

The scalar potential of the type IIB1 scenario can be derived following the same steps as the computation performed in \cite[Appendix B.3]{Marchesano:2021gyv}. We start with the standard Cremmer et al. formula \cite{Cremmer:1982en} for the F-term potential in F-theory
\begin{equation}
    e^{-K}V_F=K^{m\bar{n}}D_m W D_{\bar{n}}\bar{W}-3|W|^2\ ,
\end{equation}
where $D_m W=\partial_{m}W+(\partial_{m}K) W$, $K^{m\bar{n}}$ is the inverse field space metric and $m, n$ run over all moduli. Ignoring corrections to the K\"ahler sector of the compactification we recover the standard cancellation of no-scale structure models and the above expression simplifies to
\begin{align}
    e^{-K}V_F=&\ K^{A\bar{B}}D_{A}W D_{\bar{B}} \bar{W}\nonumber\\
    =&\ K^{A\bar{B}}\left[\Re W_A \Re W_{\bar{B}} +\Im W_A \Im W_{\bar{B}}+\left((\Re W)^2+\Im(W)^2\right)K_A\bar{K}_{\bar{B}}\right.\nonumber\\
    &\left.+K_A W\bar{W}_{\bar{B}}+\bar{K}_{\bar{B}}W_A \bar{W}\right]\ ,
\end{align}
with $W_A\equiv\partial_A W$ and now $A,B\in\{0,i\}$ only run over the dilaton and complex structure moduli.

Using our knowledge of the metric and its properties, the above expression can be expanded to
\begin{align}
    e^{-K}V_F=&\ \frac{4-2\epsilon}{1-2\epsilon}\left((\Re W)^2+(\Im W)^2\right)+K^{ij}(\Re W_i\Re W_j+ \Im W_i\Im W_j)\nonumber\\
    &+4t^i \frac{1+\epsilon}{1-2\epsilon}[\Re W \Im W_i- \Im W \Re W_i]+4(t^0)^2[(\Re W_0)^2+(\Im W_0)^2]\nonumber\\
    &+4t^0[\Re W\Im W_0-\Im W\Re W_0]\ .
\end{align}
 We proceed to consider the version of the superpotential and the Kähler potential  described in the main text in eq.~\eqref{eq:W_bilinear}:
\begin{equation}
W=\half\vec{Z}^t M \vec{Z} + \vec{L} \cdot \vec{Z} +Q\ ,
\end{equation}
Splitting the real and imaginary parts we see that
\begin{align}
\begin{split}
        \Re W=&\ \frac{1}{2}\vec{B}M\vec{B}+\vec{L}\cdot\vec{B}+Q-\frac{1}{2}\vec{T}M\vec{T}=\rho-\frac{1}{2}\kappa_i f_A^i+t^0t^ih_i^B\ ,\\
        \Im W=&\ \vec{T}\cdot(M\vec{B}+\vec{L})=\rho_A t^A\ ,
\end{split}        
\end{align}
where we have defined
\begin{align}
\begin{split}
        \rho&\equiv\frac{1}{2}\vec{B}M\vec{B}+\vec{L}\cdot\vec{B}+Q\ ,\\
        \rho_A&\equiv M_{AB}b^B+L_A\ .
\end{split}
\end{align}
Similarly, the real and imaginary parts of the partial derivatives of the superpotential can be written as follows:
\begin{eqnarray}
                \Re W_0=&\ \bar{\rho}_0\ , \qquad & \Im W_0=-t^ih_i^B\ ,\nonumber\\
        \Re W_i=&\ \bar{\rho}_i\ , \qquad & \Im W_i= -t^0h_i^B+ \kappa_{ij}f_A^j\ .
\end{eqnarray}
Substituting, expanding, rearranging and using the expressions found in the previous section we conclude that
\begin{align}
    V_Fe^{-K}=&\ 4\rho^2+4(\rho_0t^0)^2+(1+\xi)\left(K_{\rm o}^{ij}\rho_i\rho_j+(t^0)^2 K_{\rm o}^{ij} h_i^B h_j^B+\frac{4}{9}K_{ij}^{\rm o} f^i_Af^j_A+\frac{4}{3}t^0\kappa h_i^B f^i_A\right)\nonumber\\
    &+\frac{\xi}{1-2\xi}\bigg[6\rho^2+6\rho\kappa_if_A^i+\frac{1}{2}(\kappa_if_A^i)^2+6(\rho_0 t^0)^2-2(\rho_it^i)^2\!-2(t^0h_i^Bt^i)^2\\
    &+2\xi\left[2(\rho_it^i)^2+2(t^it^0h_i^B)^2+(\kappa_i f^i_A)^2\right]\bigg]\ ,\nonumber
\end{align}
which at leading order recovers the result \cite[eq. (4.18)]{Marchesano:2021gyv} in the IIB1 scenario.

\subsection{Hessian}

Now that we have the potential, we can compute the second derivatives. We focus only on the simpler axionic directions. For that mission, the following relations prove to be very useful.
\begin{equation}
\begin{aligned}
&\frac{\partial \rho}{\partial b^0}=\rho_0\ , &&\frac{\partial \rho}{\partial b^i}=\rho_i\ , &&\frac{\partial \rho_0}{\partial b^0}=0\ ,\\ 
&\frac{\partial \rho_0}{\partial b^i}=-h_i^B\ ,\qquad\qquad &&\frac{\partial\rho_i}{\partial b^0}=-h_i^B\ ,\qquad\qquad  &&\frac{\partial \rho_i}{\partial b^j}=\kappa_{ijk}f^k_A\ .
\end{aligned}
\end{equation}
Thanks to them we obtain that the first derivatives can be written as
    \begin{align}
        \frac{\partial V}{\partial b^0}e^{-K}=&\ 8\rho\rho_0-(1+\xi)(2K^{ij}_{\rm o}h_i^B\rho_j)\nonumber\\
        &+\frac{3\xi}{1-2\xi}(4\rho\rho_0+2\rho_0\kappa_i f_A^i+\frac{4}{3}h^B_it^i\rho_j t^j -\frac{8}{3}\xi h^B_i t^i\rho_jt^j)\ ,\\
        \frac{\partial V}{\partial b^i}e^{-K}=& \ 8\rho\rho_i-8\rho_0t^0h_i^B t^0+(1+\xi)(2K_{\rm o}^{jk}\kappa_{ijl}f_A^l\rho_k)\nonumber\\
        &+\frac{3\xi}{1-2\xi}\left[4\rho\rho_i+2\rho_i\kappa_jf_A^j-4h_i^B\rho_0(t^0)^2-\frac{4}{3}\rho_jt^j\kappa_{ik}f_A^k+\frac{8}{3}\xi \rho_jt^j\kappa_{ik}f_A^k\right]\ .\nonumber
    \end{align}
We proceed with the second derivatives. From the above expressions we can already see that axions and saxions are decoupled in the vacuum. Noting that  that the $\rho$'s do not depend on the saxions and that the equation of motion \eqref{eq:B_gen} implies $\rho_A=0$, it is easy to see that the cross terms involving derivatives of saxions and axions vanish. Therefore, the saxionic and axionic mass matrices are decoupled. Focusing on the pure axionic sector we find
    \begin{align}
        \frac{\partial^2 V}{(\partial b^0)^2}e^{-K}=&\ 8\rho_0^2+(1+\xi)(2K_{\rm o}^{ij}h_i^Bh_j^B)+\frac{3\xi}{1-2\xi}\left(4\rho_0^2-\frac{4}{3}(h_i^Bt^i)^2+\frac{8}{3}\xi (h_i^Bt^i)^2\right)\ ,\nonumber\\
        \frac{\partial^2 V}{\partial b^i\partial b^0}e^{-K}=&\ 8\rho_i\rho_0-8\rho h_i^B-(1+\xi)2K_{\rm o}^{jk}\kappa_{ijl}f_A^l h_k^B\\
        &+\frac{3\xi}{1-2\xi}(4\rho_i\rho_0-4\rho h_i^B-2h_i^B\kappa_jf_A^j+\frac{4}{3}h_j^Bt^j\kappa_{ik}f_A^k-\frac{8}{3}\xi h_j^Bt^j \kappa_{ik}f_A^k)\ ,\nonumber\\
         \frac{\partial^2 V}{\partial b^i\partial b^j}e^{-K}=&\ 8\rho_i\rho_j+8\rho\kappa_{ijk}f_A^k+8h_i^Bh_j^B(t^0)^2+(1+\xi)(2K^{kl}_{\rm o}\kappa_{ikm}\kappa_{jln}f_A^mf_A^n)\nonumber\\
         &+\frac{3\xi}{1-2\xi}(4\rho_i\rho_j+4\rho\kappa_{ijk}f_A^k+2\kappa_{ijk}f_A^j\kappa_l f_A^l+4h_i^Bh_j^B(t^0)^2-\frac{4}{3}\kappa_{ik}f_A^k\kappa_{jl}f_A^l\nonumber\\
         &+\frac{8}{3}\xi \kappa_{ik}\kappa_{jl}f_A^kf_A^l)\ .\nonumber
    \end{align}
To evaluate the Hessian in the vacuum, we introduce the equations of motion and restrict ourselves to the ansatz considered in the main text \eqref{eq:that}. Hence, from now on the results will be only valid in a particular subranch of the non-supersymmetric vacua with $M$ regular. The relation for the axions demands $\rho_A=0$ while the ansatz \eqref{eq:that} in combination with the equations of motion of the saxions \eqref{hBicond} implies 
\begin{equation}
    f_A^i= \frac{t^i}{\hat{t}}\ ,\qquad  h_i^B=-\frac{q^{-1}\hat{h}^B}{\hat{t}^2}\kappa_i\ .
\end{equation}
For the sake of convenience we rewrite the last two relations in terms of the coefficients of the decomposition introduced in \eqref{decomp}. Then, with the help of \eqref{eq:t0t} we have the simple relations
\begin{equation}
    f_A^i=-t^0 r_\xi B\ , \qquad h_i^B= B\kappa_i\ ,
\end{equation}
where $B=-A/(t^0 r_\xi)$, $A=1/\hat{t}$ and we have defined
\begin{equation}
    r_\xi\equiv \frac{2-\xi}{1+\xi}\ .
\end{equation}
Finally, when the axionic equations of motion are satisfied, eq.~\eqref{eq:B_gen} means $\rho=Q'$ and using \eqref{eq:flux_const_away} and the above definitions we can derive the following equation
\begin{equation}
    \rho=\frac{3}{2}\frac{\xi}{\xi+1}\kappa t^0 B\ .
\end{equation}
Putting all together, we conclude that the Hessian evaluated in the branch \eqref{eq:that} takes the form
    \begin{align}
    \label{eq: axionic derivatives + ansatz}
        \frac{\partial^2 V}{(\partial b^0)^2}e^{-K}=&\ \frac{4}{3}B^2(2-\xi)\kappa^2\ ,\nonumber\\
        \frac{\partial^2 V}{\partial b^i\partial b^0}e^{-K}=&\ \frac{4B^2\kappa t^0 (2-\xi)^2}{3(1+\xi)}\kappa_i \ ,\\
         \frac{\partial^2 V}{\partial b^i\partial b^j}e^{-K}=&\ \frac{4 B^2 \left(2 \xi^4-16 \xi^3+30 \xi^2-19 \xi+14\right) (t^0)^2}{3 (\xi+1)^2 (2 \xi-1)}\kappa_i\kappa_j+\frac{8 B^2 (\xi-2)^2 k^2 (t^0)^2}{9 (\xi+1)} K_{ij}^0\ .\nonumber
    \end{align}
The last step is to write the Hessian for canonically normalized fields. We separate the dilaton and the non-primitive directions by considering an orthogonal basis of the form $\mathcal{B}\equiv\{e_0,e_1, e_\alpha\}$ where the elements are chosen such that $e_0^0K_{00}e_0^0=1$, $e_1^i K_{ij}^{\rm NP} e_1^j=1$ and  $e_\alpha^i K_{ij}^{\rm oP}e_\alpha^j=1$ $\forall\alpha$, with $K_{ij}^{\rm P} e^i_1=0=K_{ij}^{\rm NP} e^j_{\alpha}$. To make the basis explicit we make use of \eqref{eq: primitive and no primitive corrected}. We have
    \begin{align}
    \begin{split}
        e_0=&\ \{2t^0,0,\dots, 0\}\ ,\\
        e_1^a=&\ \frac{2}{\sqrt{3}}\frac{1+\xi}{\sqrt{1-2\xi}}t^a\ .
    \end{split}
    \end{align}
Note that since $K_{ij}^{\rm P} e^i_\alpha e^j_\beta=\delta_{\alpha \beta}$, then $K_{ij}^{\rm oP} e^i_\alpha e^j_\beta=\delta_{\alpha \beta} (1+\xi)$. Projecting the  Hessian \eqref{eq: axionic derivatives + ansatz} along the directions of our canonically normalized basis, we obtain the final following form:
\begin{equation}
    H=e^K B^2\kappa^2 (t^0)^2\left(\begin{array}{ccc}
        \frac{16}{3}(2 - \xi) & \frac{16(2-\xi)^2}{3\sqrt{3-6\xi}} & 0 \\
        \frac{16(2-\xi)^2}{3\sqrt{3-6\xi}} & \frac{16 \left(2 \xi^4-16 \xi^3+30 \xi^2-19 \xi+14\right) }{9 (1-2 \xi)^2} & 0 \\
        0 & 0 & \frac{8}{9}(2-\xi)^2
    \end{array}\right)\, .
\end{equation}
Ignoring the global factors, this matrix has the following eigenvalues:
\begin{align}
\begin{split}
    \lambda_1=&\ \frac{16 (\xi-2)}{9 (2 \xi-1)^3} \left(2 \xi^4-25 \xi^3+30 \xi^2-19 \xi+5\right.\\
    &\left.+\sqrt{1-2 \xi} \sqrt{-(\xi-2)^3 \left(2 \xi^4-37 \xi^3+30 \xi^2-10 \xi+2\right)}\right)\ ,\\
    \lambda_2=&\ \frac{16 (\xi-2)}{9 (2 \xi-1)^3} \left(2 \xi^4-25 \xi^3+30 \xi^2-19 \xi+5\right.\\
    &\left.-\sqrt{1-2 \xi} \sqrt{-(\xi-2)^3 \left(2 \xi^4-37 \xi^3+30 \xi^2-10 \xi+2\right)}\right)\ ,\\
    \lambda_3=&\ \frac{8}{9} (\xi-2)^2\ .
\end{split}
\end{align} 
The first two eigenvalues have multiplicity one whereas the last one has multiplicity $h^{2,1}-1$. Adding the factors and remembering  that the mass spectrum gets and additional factor $1/2$, the masses will be given by
\begin{equation}
    m_i^2=\frac{1}{2}e^K A^2 \kappa^2 (t^0)^2 \lambda_i\, M_{\rm P}^2\ .
\end{equation}
To compare with the results found in \eqref{eq:nsa_mass_spectrum_app}, we expand the exponential of the Kähler potential
\begin{equation}
    e^K=\frac{1}{\mathcal{V}^2}\frac{1}{2t^0}\frac{1}{\frac{4}{3}\kappa(1+\xi)}\ ,
\end{equation}
and the gravitino mass
\begin{equation}
m_{3/2}^2=\frac{3}{2\mathcal{V}^2}\frac{\Nf}{2-\xi} M_{\rm P}^2  =  \frac{3}{2\mathcal{V}^2}\frac{B^2 t^0 \kappa }{1+\xi} M_{\rm P}^2 \, .
\end{equation}
Putting all together we conclude that the  eigenvalues coincide with the results in \eqref{eq:nsa_mass_spectrum_app}. This calculation has the advantage that it enables us to distinguish the axionic and saxionic masses. The axionic ones under consideration here then correspond to the following choices of signs in \eqref{eq:nsa_mass_spectrum_app}
\begin{align}
\begin{split}
    m_1^2= &\ m_{3/2}^2 \left( 1 + \sqrt{\frac{1 -2 \xi}{3}} (\hat{m} (\xi))^{-1} \right)^2\ ,\\
    m_2^2 = &\ m_{3/2}^2 \left( 1 - \sqrt{\frac{1 -2 \xi}{3}} \hat{m} (\xi) \right)^2\ ,\\
    m_3^2 =& \ m_{3/2}^2 \left(1-\frac{1+\xi}{3}\right)^2\ .
\end{split}
\end{align}

\bibliography{refs}

\providecommand{\href}[2]{#2}\begingroup\raggedright\begin{thebibliography}{10}

\bibitem{Grana:2005jc}
M.~Grana, \emph{{Flux compactifications in string theory: A Comprehensive
  review}}, \href{https://doi.org/10.1016/j.physrep.2005.10.008}{\emph{Phys.
  Rept.} {\bfseries 423} (2006) 91}
  [\href{https://arxiv.org/abs/hep-th/0509003}{{\ttfamily hep-th/0509003}}].

\bibitem{Douglas:2006es}
M.R.~Douglas and S.~Kachru, \emph{{Flux compactification}},
  \href{https://doi.org/10.1103/RevModPhys.79.733}{\emph{Rev. Mod. Phys.}
  {\bfseries 79} (2007) 733}
  [\href{https://arxiv.org/abs/hep-th/0610102}{{\ttfamily hep-th/0610102}}].

\bibitem{Blumenhagen:2006ci}
R.~Blumenhagen, B.~Kors, D.~Lust and S.~Stieberger, \emph{{Four-dimensional
  String Compactifications with D-Branes, Orientifolds and Fluxes}},
  \href{https://doi.org/10.1016/j.physrep.2007.04.003}{\emph{Phys. Rept.}
  {\bfseries 445} (2007) 1}
  [\href{https://arxiv.org/abs/hep-th/0610327}{{\ttfamily hep-th/0610327}}].

\bibitem{Becker:2006dvp}
K.~Becker, M.~Becker and J.H.~Schwarz, \emph{{String theory and M-theory: A
  modern introduction}}, Cambridge University Press (12, 2006),
  \href{https://doi.org/10.1017/CBO9780511816086}{10.1017/CBO9780511816086}.

\bibitem{Marchesano:2007de}
F.~Marchesano, \emph{{Progress in D-brane model building}},
  \href{https://doi.org/10.1002/prop.200610381}{\emph{Fortsch. Phys.}
  {\bfseries 55} (2007) 491}
  [\href{https://arxiv.org/abs/hep-th/0702094}{{\ttfamily hep-th/0702094}}].

\bibitem{Denef:2008wq}
F.~Denef, \emph{{Les Houches Lectures on Constructing String Vacua}},
  {\emph{Les Houches} {\bfseries 87} (2008) 483}
  [\href{https://arxiv.org/abs/0803.1194}{{\ttfamily 0803.1194}}].

\bibitem{Denef:2007pq}
F.~Denef, M.R.~Douglas and S.~Kachru, \emph{{Physics of String Flux
  Compactifications}},
  \href{https://doi.org/10.1146/annurev.nucl.57.090506.123042}{\emph{Ann. Rev.
  Nucl. Part. Sci.} {\bfseries 57} (2007) 119}
  [\href{https://arxiv.org/abs/hep-th/0701050}{{\ttfamily hep-th/0701050}}].

\bibitem{Ibanez:2012zz}
L.E.~Ibanez and A.M.~Uranga, \emph{{String theory and particle physics: An
  introduction to string phenomenology}}, Cambridge University Press (2, 2012).

\bibitem{Quevedo:2014xia}
F.~Quevedo, \emph{{Local String Models and Moduli Stabilisation}},
  \href{https://doi.org/10.1142/S0217732315300049}{\emph{Mod. Phys. Lett. A}
  {\bfseries 30} (2015) 1530004}
  [\href{https://arxiv.org/abs/1404.5151}{{\ttfamily 1404.5151}}].

\bibitem{Baumann:2014nda}
D.~Baumann and L.~McAllister, \emph{{Inflation and String Theory}}, Cambridge
  Monographs on Mathematical Physics, Cambridge University Press (5, 2015),
  \href{https://doi.org/10.1017/CBO9781316105733}{10.1017/CBO9781316105733},
  [\href{https://arxiv.org/abs/1404.2601}{{\ttfamily 1404.2601}}].

\bibitem{Giddings:2001yu}
S.B.~Giddings, S.~Kachru and J.~Polchinski, \emph{{Hierarchies from fluxes in
  string compactifications}},
  \href{https://doi.org/10.1103/PhysRevD.66.106006}{\emph{Phys. Rev. D}
  {\bfseries 66} (2002) 106006}
  [\href{https://arxiv.org/abs/hep-th/0105097}{{\ttfamily hep-th/0105097}}].

\bibitem{Gukov:1999ya}
S.~Gukov, C.~Vafa and E.~Witten, \emph{{CFT's from Calabi-Yau four folds}},
  \href{https://doi.org/10.1016/S0550-3213(00)00373-4}{\emph{Nucl. Phys. B}
  {\bfseries 584} (2000) 69}
  [\href{https://arxiv.org/abs/hep-th/9906070}{{\ttfamily hep-th/9906070}}].

\bibitem{Lanza:2019xxg}
S.~Lanza, F.~Marchesano, L.~Martucci and D.~Sorokin, \emph{{How many fluxes fit
  in an EFT?}}, \href{https://doi.org/10.1007/JHEP10(2019)110}{\emph{JHEP}
  {\bfseries 10} (2019) 110}
  [\href{https://arxiv.org/abs/1907.11256}{{\ttfamily 1907.11256}}].

\bibitem{Kachru:2003aw}
S.~Kachru, R.~Kallosh, A.D.~Linde and S.P.~Trivedi, \emph{{De Sitter vacua in
  string theory}},
  \href{https://doi.org/10.1103/PhysRevD.68.046005}{\emph{Phys. Rev. D}
  {\bfseries 68} (2003) 046005}
  [\href{https://arxiv.org/abs/hep-th/0301240}{{\ttfamily hep-th/0301240}}].

\bibitem{Balasubramanian:2005zx}
V.~Balasubramanian, P.~Berglund, J.P.~Conlon and F.~Quevedo, \emph{{Systematics
  of moduli stabilisation in Calabi-Yau flux compactifications}},
  \href{https://doi.org/10.1088/1126-6708/2005/03/007}{\emph{JHEP} {\bfseries
  03} (2005) 007} [\href{https://arxiv.org/abs/hep-th/0502058}{{\ttfamily
  hep-th/0502058}}].

\bibitem{Conlon:2005ki}
J.P.~Conlon, F.~Quevedo and K.~Suruliz, \emph{{Large-volume flux
  compactifications: Moduli spectrum and D3/D7 soft supersymmetry breaking}},
  \href{https://doi.org/10.1088/1126-6708/2005/08/007}{\emph{JHEP} {\bfseries
  08} (2005) 007} [\href{https://arxiv.org/abs/hep-th/0505076}{{\ttfamily
  hep-th/0505076}}].

\bibitem{Westphal:2006tn}
A.~Westphal, \emph{{de Sitter string vacua from Kahler uplifting}},
  \href{https://doi.org/10.1088/1126-6708/2007/03/102}{\emph{JHEP} {\bfseries
  03} (2007) 102} [\href{https://arxiv.org/abs/hep-th/0611332}{{\ttfamily
  hep-th/0611332}}].

\bibitem{Giryavets:2003vd}
A.~Giryavets, S.~Kachru, P.K.~Tripathy and S.P.~Trivedi, \emph{{Flux
  compactifications on Calabi-Yau threefolds}},
  \href{https://doi.org/10.1088/1126-6708/2004/04/003}{\emph{JHEP} {\bfseries
  04} (2004) 003} [\href{https://arxiv.org/abs/hep-th/0312104}{{\ttfamily
  hep-th/0312104}}].

\bibitem{Giryavets:2004zr}
A.~Giryavets, S.~Kachru and P.K.~Tripathy, \emph{{On the taxonomy of flux
  vacua}}, \href{https://doi.org/10.1088/1126-6708/2004/08/002}{\emph{JHEP}
  {\bfseries 08} (2004) 002}
  [\href{https://arxiv.org/abs/hep-th/0404243}{{\ttfamily hep-th/0404243}}].

\bibitem{DeWolfe:2004ns}
O.~DeWolfe, A.~Giryavets, S.~Kachru and W.~Taylor, \emph{{Enumerating flux
  vacua with enhanced symmetries}},
  \href{https://doi.org/10.1088/1126-6708/2005/02/037}{\emph{JHEP} {\bfseries
  02} (2005) 037} [\href{https://arxiv.org/abs/hep-th/0411061}{{\ttfamily
  hep-th/0411061}}].

\bibitem{Denef:2004dm}
F.~Denef, M.R.~Douglas and B.~Florea, \emph{{Building a better racetrack}},
  \href{https://doi.org/10.1088/1126-6708/2004/06/034}{\emph{JHEP} {\bfseries
  06} (2004) 034} [\href{https://arxiv.org/abs/hep-th/0404257}{{\ttfamily
  hep-th/0404257}}].

\bibitem{Louis:2012nb}
J.~Louis, M.~Rummel, R.~Valandro and A.~Westphal, \emph{{Building an explicit
  de Sitter}}, \href{https://doi.org/10.1007/JHEP10(2012)163}{\emph{JHEP}
  {\bfseries 10} (2012) 163} [\href{https://arxiv.org/abs/1208.3208}{{\ttfamily
  1208.3208}}].

\bibitem{Cicoli:2013cha}
M.~Cicoli, D.~Klevers, S.~Krippendorf, C.~Mayrhofer, F.~Quevedo and
  R.~Valandro, \emph{{Explicit de Sitter Flux Vacua for Global String Models
  with Chiral Matter}},
  \href{https://doi.org/10.1007/JHEP05(2014)001}{\emph{JHEP} {\bfseries 05}
  (2014) 001} [\href{https://arxiv.org/abs/1312.0014}{{\ttfamily 1312.0014}}].

\bibitem{Klemm:1992tx}
A.~Klemm and S.~Theisen, \emph{{Considerations of one modulus Calabi-Yau
  compactifications: Picard-Fuchs equations, Kahler potentials and mirror
  maps}}, \href{https://doi.org/10.1016/0550-3213(93)90289-2}{\emph{Nucl. Phys.
  B} {\bfseries 389} (1993) 153}
  [\href{https://arxiv.org/abs/hep-th/9205041}{{\ttfamily hep-th/9205041}}].

\bibitem{Doran:2007jw}
C.~Doran, B.~Greene and S.~Judes, \emph{{Families of quintic Calabi-Yau 3-folds
  with discrete symmetries}},
  \href{https://doi.org/10.1007/s00220-008-0473-x}{\emph{Commun. Math. Phys.}
  {\bfseries 280} (2008) 675}
  [\href{https://arxiv.org/abs/hep-th/0701206}{{\ttfamily hep-th/0701206}}].

\bibitem{Candelas:2017ive}
P.~Candelas and C.~Mishra, \emph{{Highly Symmetric Quintic Quotients}},
  \href{https://doi.org/10.1002/prop.201800017}{\emph{Fortsch. Phys.}
  {\bfseries 66} (2018) 1800017}
  [\href{https://arxiv.org/abs/1709.01081}{{\ttfamily 1709.01081}}].

\bibitem{Braun:2011hd}
V.~Braun, \emph{{The 24-Cell and Calabi-Yau Threefolds with Hodge Numbers
  (1,1)}}, \href{https://doi.org/10.1007/JHEP05(2012)101}{\emph{JHEP}
  {\bfseries 05} (2012) 101} [\href{https://arxiv.org/abs/1102.4880}{{\ttfamily
  1102.4880}}].

\bibitem{Batyrev:2008rp}
V.~Batyrev and M.~Kreuzer, \emph{{Constructing new Calabi-Yau 3-folds and their
  mirrors via conifold transitions}},
  \href{https://doi.org/10.4310/ATMP.2010.v14.n3.a3}{\emph{Adv. Theor. Math.
  Phys.} {\bfseries 14} (2010) 879}
  [\href{https://arxiv.org/abs/0802.3376}{{\ttfamily 0802.3376}}].

\bibitem{Doran:2005gu}
C.F.~Doran and J.W.~Morgan, \emph{{Mirror symmetry and integral variations of
  Hodge structure underlying one parameter families of Calabi-Yau threefolds}},
   in \emph{{Workshop on Calabi-Yau Varieties and Mirror Symmetry}},
  pp.~517--537, 5, 2005 [\href{https://arxiv.org/abs/math/0505272}{{\ttfamily
  math/0505272}}].

\bibitem{Candelas:2019llw}
P.~Candelas, X.~de~la Ossa, M.~Elmi and D.~Van~Straten, \emph{{A One Parameter
  Family of Calabi-Yau Manifolds with Attractor Points of Rank Two}},
  \href{https://doi.org/10.1007/JHEP10(2020)202}{\emph{JHEP} {\bfseries 10}
  (2020) 202} [\href{https://arxiv.org/abs/1912.06146}{{\ttfamily
  1912.06146}}].

\bibitem{Joshi:2019nzi}
A.~Joshi and A.~Klemm, \emph{{Swampland Distance Conjecture for One-Parameter
  Calabi-Yau Threefolds}},
  \href{https://doi.org/10.1007/JHEP08(2019)086}{\emph{JHEP} {\bfseries 08}
  (2019) 086} [\href{https://arxiv.org/abs/1903.00596}{{\ttfamily
  1903.00596}}].

\bibitem{Grimm:2019ixq}
T.W.~Grimm, C.~Li and I.~Valenzuela, \emph{{Asymptotic Flux Compactifications
  and the Swampland}},
  \href{https://doi.org/10.1007/JHEP06(2020)009}{\emph{JHEP} {\bfseries 06}
  (2020) 009} [\href{https://arxiv.org/abs/1910.09549}{{\ttfamily
  1910.09549}}].

\bibitem{Blanco-Pillado:2020wjn}
J.J.~Blanco-Pillado, K.~Sousa, M.A.~Urkiola and J.M.~Wachter, \emph{{Towards a
  complete mass spectrum of type-IIB flux vacua at large complex structure}},
  \href{https://doi.org/10.1007/JHEP04(2021)149}{\emph{JHEP} {\bfseries 04}
  (2021) 149} [\href{https://arxiv.org/abs/2007.10381}{{\ttfamily
  2007.10381}}].

\bibitem{Blanco-Pillado:2020hbw}
J.J.~Blanco-Pillado, K.~Sousa, M.A.~Urkiola and J.M.~Wachter, \emph{{Universal
  Class of Type-IIB Flux Vacua with Analytic Mass Spectrum}},
  \href{https://doi.org/10.1103/PhysRevD.103.106006}{\emph{Phys. Rev. D}
  {\bfseries 103} (2021) 106006}
  [\href{https://arxiv.org/abs/2011.13953}{{\ttfamily 2011.13953}}].

\bibitem{Marchesano:2021gyv}
F.~Marchesano, D.~Prieto and M.~Wiesner, \emph{{F-theory flux vacua at large
  complex structure}},
  \href{https://doi.org/10.1007/JHEP08(2021)077}{\emph{JHEP} {\bfseries 08}
  (2021) 077} [\href{https://arxiv.org/abs/2105.09326}{{\ttfamily
  2105.09326}}].

\bibitem{Sousa:2014qza}
K.~Sousa and P.~Ortiz, \emph{{Perturbative Stability along the Supersymmetric
  Directions of the Landscape}},
  \href{https://doi.org/10.1088/1475-7516/2015/02/017}{\emph{JCAP} {\bfseries
  02} (2015) 017} [\href{https://arxiv.org/abs/1408.6521}{{\ttfamily
  1408.6521}}].

\bibitem{Marsh:2015zoa}
M.C.D.~Marsh and K.~Sousa, \emph{{Universal Properties of Type IIB and F-theory
  Flux Compactifications at Large Complex Structure}},
  \href{https://doi.org/10.1007/JHEP03(2016)064}{\emph{JHEP} {\bfseries 03}
  (2016) 064} [\href{https://arxiv.org/abs/1512.08549}{{\ttfamily
  1512.08549}}].

\bibitem{Brodie:2015kza}
C.~Brodie and M.C.D.~Marsh, \emph{{The Spectra of Type IIB Flux
  Compactifications at Large Complex Structure}},
  \href{https://doi.org/10.1007/JHEP01(2016)037}{\emph{JHEP} {\bfseries 01}
  (2016) 037} [\href{https://arxiv.org/abs/1509.06761}{{\ttfamily
  1509.06761}}].

\bibitem{Mayr:2000as}
P.~Mayr, \emph{{Phases of supersymmetric D-branes on Kahler manifolds and the
  McKay correspondence}},
  \href{https://doi.org/10.1088/1126-6708/2001/01/018}{\emph{JHEP} {\bfseries
  01} (2001) 018} [\href{https://arxiv.org/abs/hep-th/0010223}{{\ttfamily
  hep-th/0010223}}].

\bibitem{Bilinear1}
S.~Bielleman, L.E.~Ibanez and I.~Valenzuela, \emph{{Minkowski 3-forms, Flux
  String Vacua, Axion Stability and Naturalness}},
  \href{https://doi.org/10.1007/JHEP12(2015)119}{\emph{JHEP} {\bfseries 12}
  (2015) 119} [\href{https://arxiv.org/abs/1507.06793}{{\ttfamily
  1507.06793}}].

\bibitem{Bilinear2}
F.~Carta, F.~Marchesano, W.~Staessens and G.~Zoccarato, \emph{{Open string
  multi-branched and K\"ahler potentials}},
  \href{https://doi.org/10.1007/JHEP09(2016)062}{\emph{JHEP} {\bfseries 09}
  (2016) 062} [\href{https://arxiv.org/abs/1606.00508}{{\ttfamily
  1606.00508}}].

\bibitem{Bilinear3}
A.~Herraez, L.E.~Ibanez, F.~Marchesano and G.~Zoccarato, \emph{{The Type IIA
  Flux Potential, 4-forms and Freed-Witten anomalies}},
  \href{https://doi.org/10.1007/JHEP09(2018)018}{\emph{JHEP} {\bfseries 09}
  (2018) 018} [\href{https://arxiv.org/abs/1802.05771}{{\ttfamily
  1802.05771}}].

\bibitem{Bilinear4}
F.~Marchesano, D.~Prieto, J.~Quirant and P.~Shukla, \emph{{Systematics of Type
  IIA moduli stabilisation}},
  \href{https://doi.org/10.1007/JHEP11(2020)113}{\emph{JHEP} {\bfseries 11}
  (2020) 113} [\href{https://arxiv.org/abs/2007.00672}{{\ttfamily
  2007.00672}}].

\bibitem{Bilinear5}
D.~Escobar, F.~Marchesano and W.~Staessens, \emph{{Type IIA Flux Vacua with
  Mobile D6-branes}},
  \href{https://doi.org/10.1007/JHEP01(2019)096}{\emph{JHEP} {\bfseries 01}
  (2019) 096} [\href{https://arxiv.org/abs/1811.09282}{{\ttfamily
  1811.09282}}].

\bibitem{Bilinear6}
D.~Escobar, F.~Marchesano and W.~Staessens, \emph{{Type IIA flux vacua and
  $\alpha'$-corrections}},
  \href{https://doi.org/10.1007/JHEP06(2019)129}{\emph{JHEP} {\bfseries 06}
  (2019) 129} [\href{https://arxiv.org/abs/1812.08735}{{\ttfamily
  1812.08735}}].

\bibitem{Bilinear7}
F.~Marchesano and J.~Quirant, \emph{{A Landscape of AdS Flux Vacua}},
  \href{https://doi.org/10.1007/JHEP12(2019)110}{\emph{JHEP} {\bfseries 12}
  (2019) 110} [\href{https://arxiv.org/abs/1908.11386}{{\ttfamily
  1908.11386}}].

\bibitem{Denef:2004ze}
F.~Denef and M.R.~Douglas, \emph{{Distributions of flux vacua}},
  \href{https://doi.org/10.1088/1126-6708/2004/05/072}{\emph{JHEP} {\bfseries
  05} (2004) 072} [\href{https://arxiv.org/abs/hep-th/0404116}{{\ttfamily
  hep-th/0404116}}].

\bibitem{Demirtas:2019sip}
M.~Demirtas, M.~Kim, L.~Mcallister and J.~Moritz, \emph{{Vacua with Small Flux
  Superpotential}},
  \href{https://doi.org/10.1103/PhysRevLett.124.211603}{\emph{Phys. Rev. Lett.}
  {\bfseries 124} (2020) 211603}
  [\href{https://arxiv.org/abs/1912.10047}{{\ttfamily 1912.10047}}].

\bibitem{Demirtas:2020ffz}
M.~Demirtas, M.~Kim, L.~McAllister and J.~Moritz, \emph{{Conifold Vacua with
  Small Flux Superpotential}},
  \href{https://doi.org/10.1002/prop.202000085}{\emph{Fortsch. Phys.}
  {\bfseries 68} (2020) 2000085}
  [\href{https://arxiv.org/abs/2009.03312}{{\ttfamily 2009.03312}}].

\bibitem{Cicoli:2022vny}
M.~Cicoli, M.~Licheri, R.~Mahanta and A.~Maharana, \emph{{Flux Vacua with
  Approximate Flat Directions}},
  \href{https://arxiv.org/abs/2209.02720}{{\ttfamily 2209.02720}}.

\bibitem{Candelas:1994hw}
P.~Candelas, A.~Font, S.H.~Katz and D.R.~Morrison, \emph{{Mirror symmetry for
  two parameter models. 2.}},
  \href{https://doi.org/10.1016/0550-3213(94)90155-4}{\emph{Nucl. Phys. B}
  {\bfseries 429} (1994) 626}
  [\href{https://arxiv.org/abs/hep-th/9403187}{{\ttfamily hep-th/9403187}}].

\bibitem{AbdusSalam:2020ywo}
S.~AbdusSalam, S.~Abel, M.~Cicoli, F.~Quevedo and P.~Shukla, \emph{{A
  systematic approach to K\"ahler moduli stabilisation}},
  \href{https://doi.org/10.1007/JHEP08(2020)047}{\emph{JHEP} {\bfseries 08}
  (2020) 047} [\href{https://arxiv.org/abs/2005.11329}{{\ttfamily
  2005.11329}}].

\bibitem{Bena:2020xrh}
I.~Bena, J.~Bl\r{a}b\"ack, M.~Gra\~na and S.~L\"ust, \emph{{The tadpole
  problem}}, \href{https://doi.org/10.1007/JHEP11(2021)223}{\emph{JHEP}
  {\bfseries 11} (2021) 223}
  [\href{https://arxiv.org/abs/2010.10519}{{\ttfamily 2010.10519}}].

\bibitem{Bena:2021wyr}
I.~Bena, J.~Bl\r{a}b\"ack, M.~Gra\~na and S.~L\"ust, \emph{{Algorithmically
  Solving the Tadpole Problem}},
  \href{https://doi.org/10.1007/s00006-021-01189-6}{\emph{Adv. Appl. Clifford
  Algebras} {\bfseries 32} (2022) 7}
  [\href{https://arxiv.org/abs/2103.03250}{{\ttfamily 2103.03250}}].

\bibitem{Plauschinn:2021hkp}
E.~Plauschinn, \emph{{The tadpole conjecture at large complex-structure}},
  \href{https://doi.org/10.1007/JHEP02(2022)206}{\emph{JHEP} {\bfseries 02}
  (2022) 206} [\href{https://arxiv.org/abs/2109.00029}{{\ttfamily
  2109.00029}}].

\bibitem{Lust:2021xds}
S.~L\"ust, \emph{{Large complex structure flux vacua of IIB and the Tadpole
  Conjecture}},  \href{https://arxiv.org/abs/2109.05033}{{\ttfamily
  2109.05033}}.

\bibitem{Grimm:2021ckh}
T.W.~Grimm, E.~Plauschinn and D.~van~de Heisteeg, \emph{{Moduli stabilization
  in asymptotic flux compactifications}},
  \href{https://doi.org/10.1007/JHEP03(2022)117}{\emph{JHEP} {\bfseries 03}
  (2022) 117} [\href{https://arxiv.org/abs/2110.05511}{{\ttfamily
  2110.05511}}].

\bibitem{Grana:2022dfw}
M.~Gra\~na, T.W.~Grimm, D.~van~de Heisteeg, A.~Herraez and E.~Plauschinn,
  \emph{{The tadpole conjecture in asymptotic limits}},
  \href{https://doi.org/10.1007/JHEP08(2022)237}{\emph{JHEP} {\bfseries 08}
  (2022) 237} [\href{https://arxiv.org/abs/2204.05331}{{\ttfamily
  2204.05331}}].

\bibitem{Lust:2022mhk}
S.~L\"ust and M.~Wiesner, \emph{{The Tadpole Conjecture in the Interior of
  Moduli Space}},  \href{https://arxiv.org/abs/2211.05128}{{\ttfamily
  2211.05128}}.

\bibitem{Candelas:1990pi}
P.~Candelas and X.~de~la Ossa, \emph{{Moduli Space of {Calabi-Yau} Manifolds}},
  \href{https://doi.org/10.1016/0550-3213(91)90122-E}{\emph{Nucl. Phys. B}
  {\bfseries 355} (1991) 455}.

\bibitem{Cremmer:1982en}
E.~Cremmer, S.~Ferrara, L.~Girardello and A.~Van~Proeyen, \emph{{Yang-Mills
  Theories with Local Supersymmetry: Lagrangian, Transformation Laws and
  SuperHiggs Effect}},
  \href{https://doi.org/10.1016/0550-3213(83)90679-X}{\emph{Nucl. Phys. B}
  {\bfseries 212} (1983) 413}.

\end{thebibliography}\endgroup

\end{document}